\documentclass[9.5pt,journal,compsoc]{IEEEtran}

\usepackage[utf8]{inputenc}
\usepackage{epsfig}
\usepackage{graphicx}
\usepackage{amsmath}
\usepackage{amssymb}
\usepackage{booktabs}
\usepackage[all]{xy}
\usepackage{tcolorbox}
\usepackage{tikz-cd}
\usepackage[linesnumbered,lined,boxed]{algorithm2e}

\ifCLASSOPTIONcompsoc
  \usepackage[nocompress]{cite}
\else
  \usepackage{cite}
\fi

\def\argmin{\mathop{\rm argmin}}
\def\arginf{\mathop{\rm arginf}}

\newcommand{\real}{\ensuremath{\mathbb{R}}}

\newcommand{\ltwo}{\ensuremath{\mathbb{L}^2}}

\def\argmin{\mathop{\rm argmin}}

\def \ltwo {\mathbb{L}^2}

\newtheorem{definition}{Definition}

\newtheorem{thm}{Theorem}
\newtheorem{lemma}{Lemma}

\begin{document}
\tabcolsep 1pt
\title{Shape Analysis of Functional Data with Elastic Partial Matching}
\author{Darshan~Bryner
        and~Anuj~Srivastava,~\IEEEmembership{Fellow,~IEEE}
\IEEEcompsocitemizethanks{\IEEEcompsocthanksitem D. Bryner is with Naval Surface Warfare Center Panama City Division, Panama City, FL. \protect\\
E-mail: darshan.bryner@navy.mil
\IEEEcompsocthanksitem A. Srivastava is with Department of Statistics, Florida State University, Tallahassee, FL.}
\thanks{Manuscript received February, 2021; revised XXX.}}


\markboth{Journal of \LaTeX\ Class Files,~Vol.~14, No.~8, August~2015}%
{Shell \MakeLowercase{\textit{et al.}}: Bare Demo of IEEEtran.cls for Computer Society Journals}

\IEEEtitleabstractindextext{%
\begin{abstract}
Elastic Riemannian metrics have been used successfully in the past for statistical treatments of functional and curve shape data. However, this usage has suffered from an important restriction: the function boundaries are assumed fixed and matched. Functional data exhibiting unmatched boundaries typically arise from dynamical systems with variable evolution rates such as COVID-19 infection rate curves associated with different geographical regions. In this case, it is more natural to model such data with sliding boundaries and use partial matching, i.e.,\ only a part of a function is matched to another function. Here, we develop a comprehensive Riemannian framework that allows for partial matching, comparing, and clustering of functions under both phase variability and uncertain boundaries. We extend past work by: (1) Forming a joint action of the time-warping and time-scaling groups; (2) Introducing a metric that is invariant to this joint action, allowing for a gradient-based approach to elastic partial matching; and (3) Presenting a modification that, while losing the metric property, allows one to control relative influence of the two groups. This framework is illustrated for registering and clustering shapes of COVID-19 rate curves, identifying essential patterns, minimizing mismatch errors, and reducing variability within clusters compared to previous methods.
\end{abstract}

}

\maketitle

\section{Introduction}
The field of functional data analysis (FDA) has seen a tremendous growth and activity over the last few years \cite{ramsay-silverman-2005,Ferraty2006-sp,srivastava-klassen:2016,hsing-eubank:2015,fda-reimharr}. This phenomenal interest in FDA stems in part from our growing ability to record, store, and transmit data streams that are indexed over near-continuous times. A number of books and papers, including the ones mentioned above, have highlighted clear benefits of analyzing these data streams as continuous functions rather than as discrete time-series. In FDA, one treats observed functions as elements of a function space, endows a metric structure on this space, and uses the geometry of this metric space to perform statistical analyses. The main challenges in the analysis of functional data come from the infinite dimensionality of function spaces, the contamination of data due to noise, and the presence of phase variability within the data. We highlight the last, and arguably most important, of these issues next. 
\\

\noindent {\bf Data with Phase Variability}:
Functional data displays an interesting phenomenon that makes it unique from the perspective of statistical analysis. Real life functional data often comes with phase variability, {\it i.e.}, functions are often observed with perturbations or warpings of the time axis, resulting in the horizontal movements of peaks and valleys. In other words, noise in real data manifests itself as not only vertical (or additive) but also horizontal (or compositional), reflecting an inherent lack of temporal synchronization across functions. For example, in the case of COVID-19 infection rates, the high and low points for different regions occur at different times, implying different evolution rates of the virus cycle. Several papers \cite{ramsay-li-RSSB:98,marron-etal-EJS:2014,srivastava-etal-function:2011,marron-etal-EJS:2015,Takagishi-Yadohisa} and a book \cite{srivastava-klassen:2016} have documented and formally developed the concept of phase variability in functional data with the guidance that ignoring this phase variability leads to inflated variance in the data, loss of structures, and lack of power in hypothesis testing. Indeed these papers provide ways of separating phase and amplitude components (also termed alignment or registration of functions) and then performing either individualized or joint statistical analyses
\cite{ramsay-li-RSSB:98,kneip-ramsay:2008,srivastava-klassen:2016,tucker-et-al:2013,jung-joint-PA}. The proposed methods (for phase-amplitude separation) differ in their choices of metrics, tools for optimizations, and their definitions of phase; however, all of these papers share a fundamental shortcoming. They assume that the functions are fully observed over a common interval and moreover that the endpoints match perfectly across observations.

In more mathematical terms, let $\{f_i: [0,T] \to \real, i= 1,2,\dots,n\}$ be the set of observed functions on an interval $[0,T]$. Isolating phase variability implies finding time-warping functions $\{\gamma_i\}$ such that the time-warped functions $\{ f_i \circ \gamma_i\}$ are {\it matched} (or {\it aligned}, or {\it registered}) where, in this context, a good matching generally refers to having functions with peaks and valleys co-located across observations. Here, the warping functions $\{\gamma_i\}$ are constrained to be diffeopmorphsisms of $[0,T]$ to itself such that $\gamma_i(0) = 0$ and $\gamma_i(T) = T$. This last property of time-warping functions implies that the endpoints of the data, $\{f_i(0)\}$ and $\{ f_i(T)\}$, are assumed to be already matched across all functions. 
\\

\noindent {\bf Data with Sliding Right Boundary}: In many situations where the phase variability is present not just in the interior of $[0,T]$ but also on the boundaries, the boundary matching assumption often breaks down and presents a major limitation to the current analysis. Sometimes one can assume that the left boundaries $\{f_i(0)\}$  are either synchronized or can be synchronized perhaps through a simple shift. However, the right boundaries are still variable, {\it i.e.}, $\gamma_i(T) \neq T$. This situation arises, for example, in COVID-19 data where at the end of any observed interval, the infection rate curves for different regions are at different states of evolution. Another example of uncertain boundaries arises in censored data. 
Censoring is a process of randomly truncating the observation interval before the scheduled end is reached. In mathematical terms, such observations with uncertain right boundaries are given by $\{f_i: [0,T_i] \to \real, i= 1,2,\dots,n\}$ with random end times $T_i \in \real_+$ for each observation. There is a significant literature for analyzing right-censored functional data \cite{Delaigle-Hall-2013,Belkis-2018}, especially in survival analysis \cite{survival-klein,dehan-2018}, where the aim is to model censoring times as random variables and use the distribution of $min(T, T_i)$ explicitly in the likelihood functions. However, the literature on censored data analysis does not consider the problem of phase variability and assumes that the functional data is fully registered at all time points as acquired. For this reason, the analysis naturally deteriorates when phase variability is present in the actual data.  
A commonly used, albeit naive, solution is to simply time scale (linearly) each observation so that they all have the same domain $[0,T]$. However, this fails to address some deeper issues present in the data. 
\\

\noindent {\bf Data with Both Phase Variability and Sliding Right Boundary}:
In this paper we are concerned with the situation where each observation in the data set has the following properties: (1) It contains random phase variability; and (2) It has a right endpoint located at a random time $T_i$. The challenge here is to separate the phase from the amplitude when the right endpoints are no longer synchronized. In other words, we wish to align or register functions, essentially by matching their peaks and valleys inside the domain, while simultaneously placing their floating right boundaries correctly with respect to each other. Some researchers refer to this problem as that of {\it elastic partial matching} \cite{robinson-thesis,cui-2009,elmi-2009}. 
An example of this situation arises in the analysis of Berkeley growth curves~\cite{ramsay-gasser-94}, where the start time of growth is birth but the end of the fixed 20 year observation period may fall at a different point of the growth cycle for different subjects due to individuals' differing biological clocks. Another example, mentioned earlier, involves incidence rates of the COVID-19 virus for different geographical locations where different cities/states/countries exhibit different rates of evolution in their incidence rates. The start time of the incidence can be assumed to be synchronized to the time of the first positive case, but, due to epidemiological and demographic factors, the pandemic evolution is certainly not well synchronized across regions. Different regions experience different growths and decline rates, and one needs to perform full registration to understand these patterns.

\begin{figure}[ht]
    \centering
    \begin{tabular}{cc}
    \includegraphics[height=1.2in]{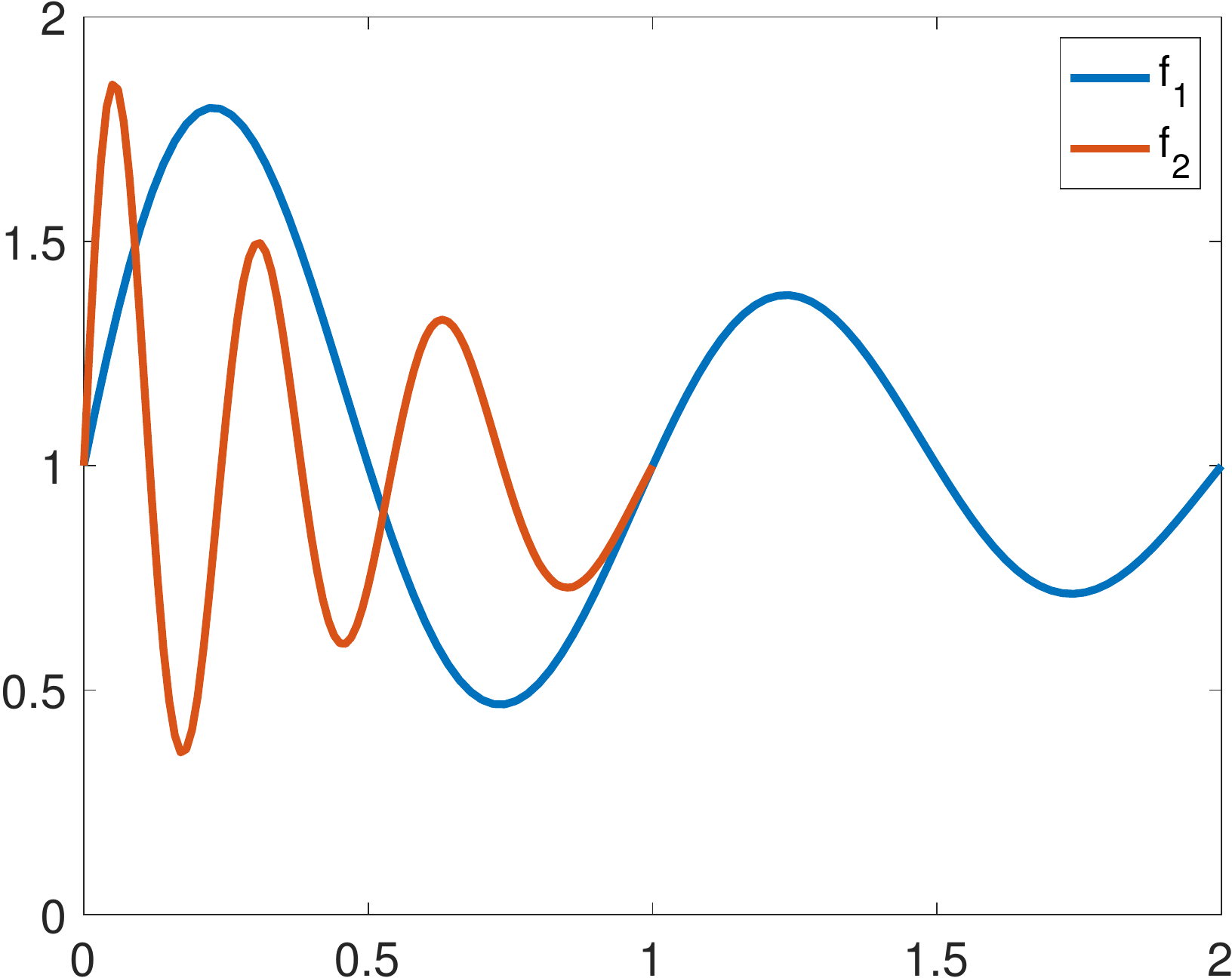} &
    \includegraphics[height=1.2in]{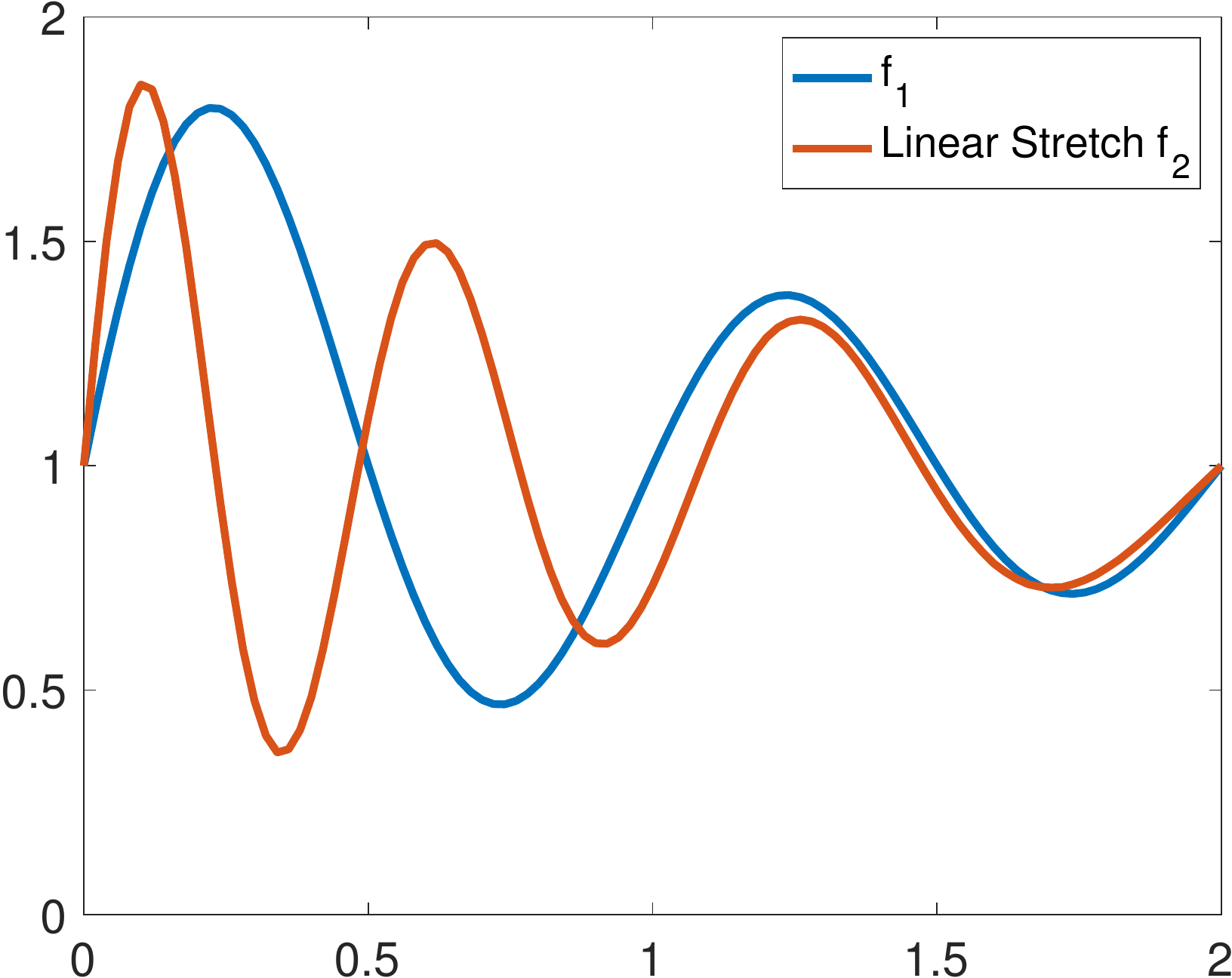} \\
    (a) Original $f_1, f_2$& (b) Linear stretch $f_2$\\
     & to match boundaries \\
    \includegraphics[height=1.2in]{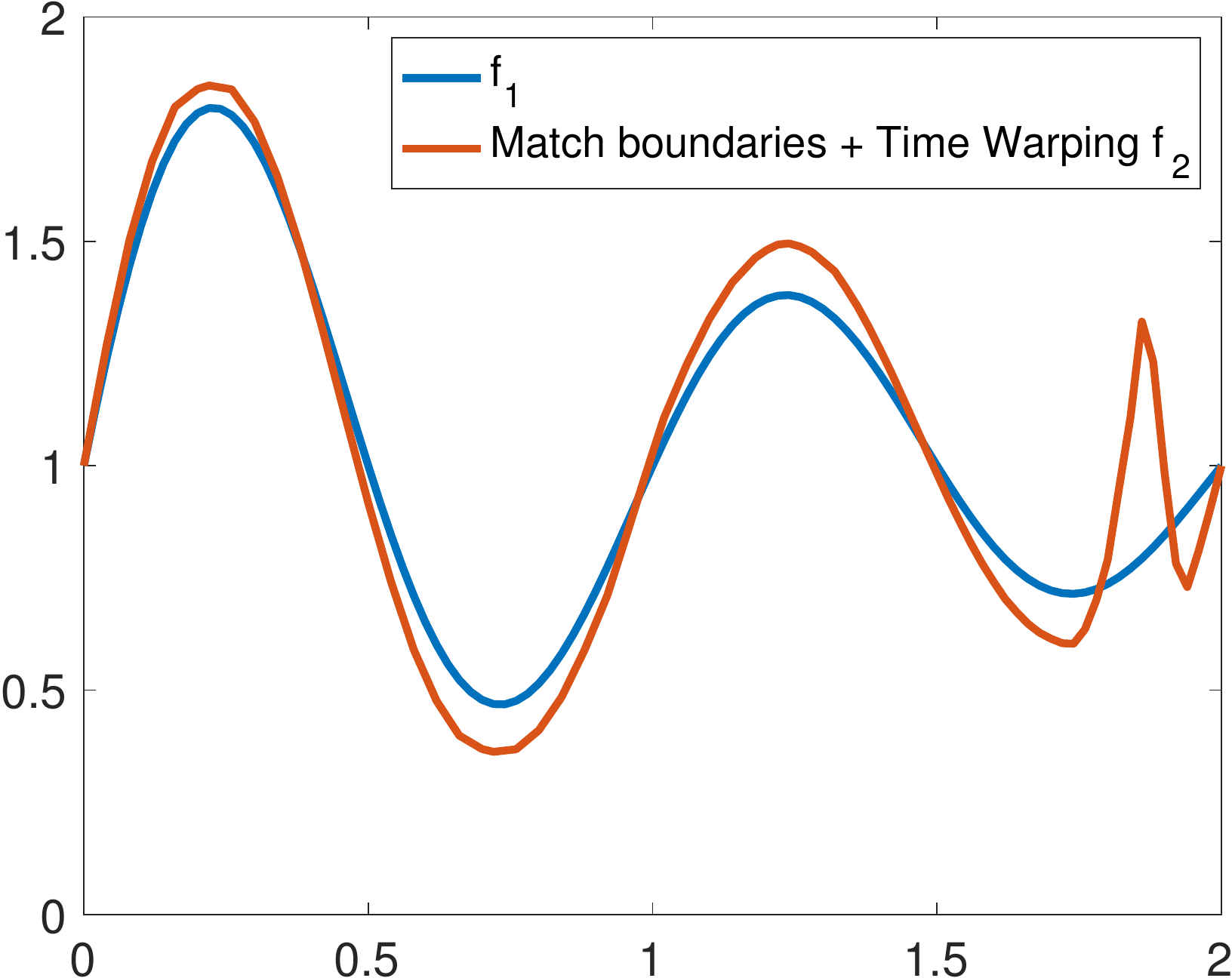} &
    \includegraphics[height=1.2in]{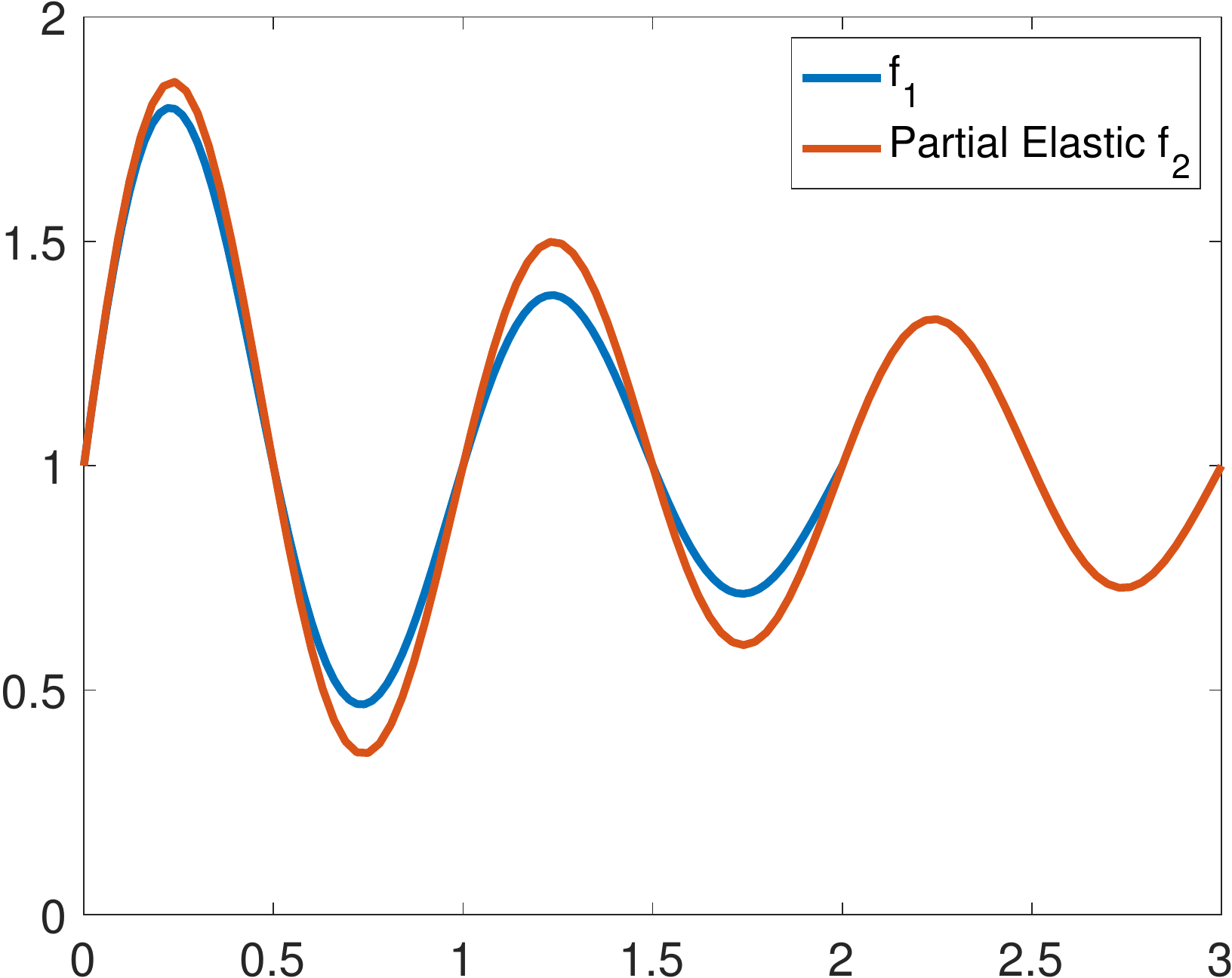} \\
    (c) Nonlinear Warping $f_2$ & (d) Warping $f_2$ \& stretching \\
     after $(b)$ & with flexible right boundary \\
    \end{tabular}
    \caption{Illustration of different potential solutions for matching curves with different shapes and flexible right boundaries.}
    \label{fig:example_data}
\end{figure}

How can we analyze such data and still preserve data structures?
We illustrate this problem with a simple example. Figure~\ref{fig:example_data}(a) shows a pair of functions, $f_1$ and $f_2$, where the left boundaries are matched well but the right boundaries are different with $T_1 = 2$ and $T_2 = 1$. Furthermore, the two functions have similar shapes except that $f_1$ is missing a piece relative to $f_2$; thus, $f_1$ can only be well matched with a part of $f_2$, despite $T_2 < T_1$.  
If we simply linearly stretch $f_2$ to match their boundaries by time scaling ($t \mapsto 2t$), we obtain the result in (b). If, in addition to their boundaries being matched by linear scaling, we also time warp $f_2$ to match with $f_1$, we get (c). Finally, if we stretch $f_2$ in such a way that $f_1$ is matched to a {\it part} of $f_2$ ($t \mapsto 3t$) and then apply time warping, we get (d). Amongst all these solutions, panel (d) provides the most  satisfactory result in matching $f_1$ and $f_2$. 

In order to reach this solution, we need to infer two items: (1) which parts of the two functions match, \textit{i.e.} how much time-scaling is needed, and (2) what is the time-warping that aligns the two matched parts? If we have a set of functions, each with an uncertain right boundary, then the joint registration, modeling, and analysis of this data becomes even more challenging. 

The following items summarize the main challenges in developing a mathematical solutions: 
\begin{enumerate}
    \item {\bf Mathematical Representations \& Metrics}: One needs mathematical representations that can account for uncertain boundaries and time-warpings of data. Specifically, we need metrics and objective functions such that partial matching of functions can be posed as binary optimization problems. Similar to past works on elastic shape analysis of functions~\cite{srivastava-klassen:2016}, we need metrics that are invariant to actions of nuisance groups applicable here -- time-warping and time-scaling.
    \item {\bf Interaction of Warping \& Sliding Boundaries}: While one has solutions for the individual problems -- time-warping to match extrema and time-scaling to match the boundaries -- the combination of the two makes the problem much more difficult. The combination requires diffeomorphic transformations of domains but without a fixed right boundary. This, in turn, demands searching over all combinations of linear stretches and nonlinear warpings to match any two functions. One needs to exploit the geometries of these two variables to help optimize over the joint space in an efficient manner. 
\end{enumerate}

This formulation resembles the problem of partial matching of shapes and there are
a number of papers in the literature on this topic,
see \cite{shilane:2006,mahesh:2011,partial-shape:2018,Alt95,buchin09,McBride03} for example. However, our focus is on Riemannian approaches as they provide a comprehensive toolbox for statistics analysis of shapes, including geodesics, proper metrics, statistical summaries, and probabilistic modeling.
In this paper we develop a comprehensive Riemannian framework for elastic analysis of functional data with right floating boundaries. Similar to the past works, we develop a framework where we represent both time-warping and time-scaling as group actions on a set of functions. Any two functions are compared by optimally time-warping their domains and varying the right boundaries using a fully automated optimization procedure. The key contribution here is a metric that is invariant to the joint action of these groups and that facilitates partial matching of curves. Additionally, we include an optional modification of this metric formulation that allows one to choose a scalar weight $\lambda$ that balances the dual goal of matching the interiors and matching the right boundaries; however, some of the metric properties are lost for values $\lambda\neq 1$. 

The rest of this paper is organized as follows. We develop the mathematical framework for partially aligning functions using an elastic Riemannian metric and a square-root representation in Section 2. The process of optimizing an objective function for pairwise matching of functions is developed in Section 3. Section 4 presents a set of experimental results involving both simulated and real data (COVID-19 infection rate curves) to demonstrate the success of the proposed framework. The paper ends with a short summary and some conclusions in Section 5.

\section{Proposed Mathematical Framework}
In this section, we develop an elastic Riemannian framework for representing, partially matching, and comparing functional data with phase variability and sliding right boundaries. Before we develop our approach, we summarize the past ideas for phase-amplitude separation, or registration of functions with matched boundaries.
We will follow the approach presented in \cite{srivastava-klassen:2016} and refer the reader to that book for more details. 

\subsection{Past Work in the Alignment of Functions with Fixed Boundaries}
\label{sec:past-work}
Let ${\cal F} = \{ f: [0,T] \to \real| \mbox{$f$ is absolutely continuous}\}$ be the set of functions of interest. Let $\Gamma_T$ be the group of all diffeomorphisms from $[0,T]$ to itself that preserve the boundaries. For any $f \in {\cal F}$ and $\gamma \in\Gamma_T$, the composition $f \circ \gamma$ denotes a time warping of $f$ while keeping the boundaries fixed ($\gamma(0) = 0$, $\gamma(T) = T$). In order to register functions,  one represents them by their square-root velocity functions (SRVFs). For any $f \in {\cal F}$, its SRVF is given by $q(t)= \mbox{sign}(\dot{f}(t)) \sqrt{|\dot{f}(t)|}$. 
In fact, this mapping $f \mapsto q$ forms a bijection between ${\cal F}$ and $\ltwo([0,T],\real)$, up to a constant.
One can reconstruct a function $f$ from its SRVF $q$ and $f(0)$ using
$f(t) = f(0) + \int_0^t q(s)|q(s)|~ds$, $t \in [0,T]$.

For any $\gamma \in \Gamma_T$, the SRVF of the time-warped function $f \circ \gamma$ is given by $(q \odot \gamma) \overset{\Delta}{=} (q \circ \gamma) \sqrt{\dot{\gamma}}$. A very important property of SRVFs is that for any two functions
$f_1, f_2 \in {\cal F}$, and their SRVFs $q_1, q_2$, we have $\|q_1 - q_2\|= \| (q_1 \odot \gamma) - (q_2 \odot \gamma)\|$ for all $\gamma \in \Gamma_T$. (The norm here is the standard $\ltwo$ norm: $\|q\|= \sqrt{\int_0^T q(t)^2~dt}$.) Due to this property, the $\ltwo$ norm between SRVFs is called the {\it elastic metric}. With the elastic metric, the problem of registration between functions can be solved as the following optimization: 
\begin{equation}
\gamma^* = \argmin_{\gamma \in \Gamma_T} \| q_1 - (q_2 \odot \gamma)\|\ .
\label{eqn:elastic-dist}
\end{equation}
This optimization is solved either using the dynamic programming algorithm or a gradient-based approach, as needed, and the resulting registration is called {\it dense elastic registration}. For any $t \in [0,T]$, the point $f_1(t)$ is said to be matched to the point $f_2(\gamma^*(t))$.

\begin{figure}
    \centering
    \begin{tabular}{ccc}
    \includegraphics[height=1.in]{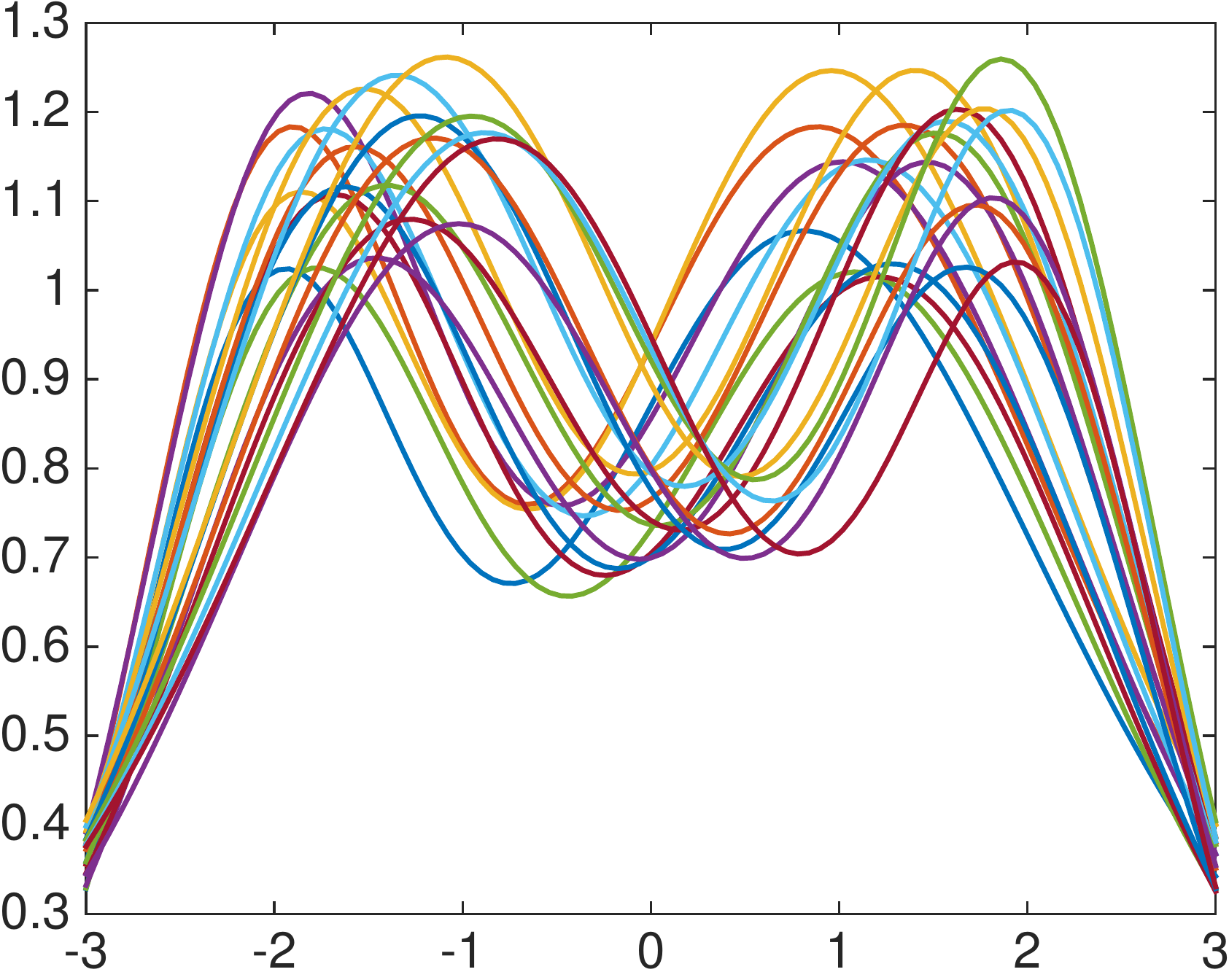} &
     \includegraphics[height=1.in]{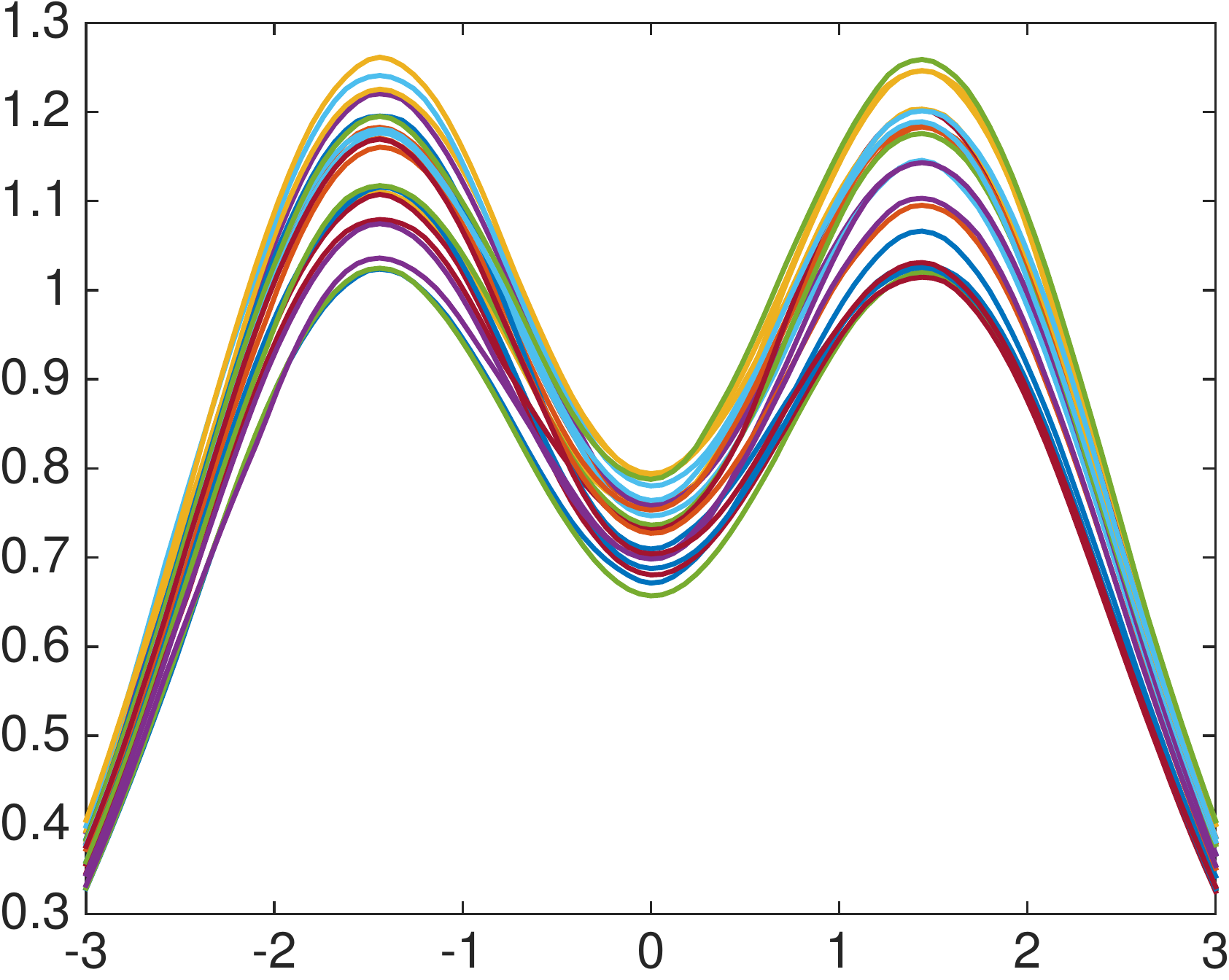} &
      \includegraphics[height=1.in]{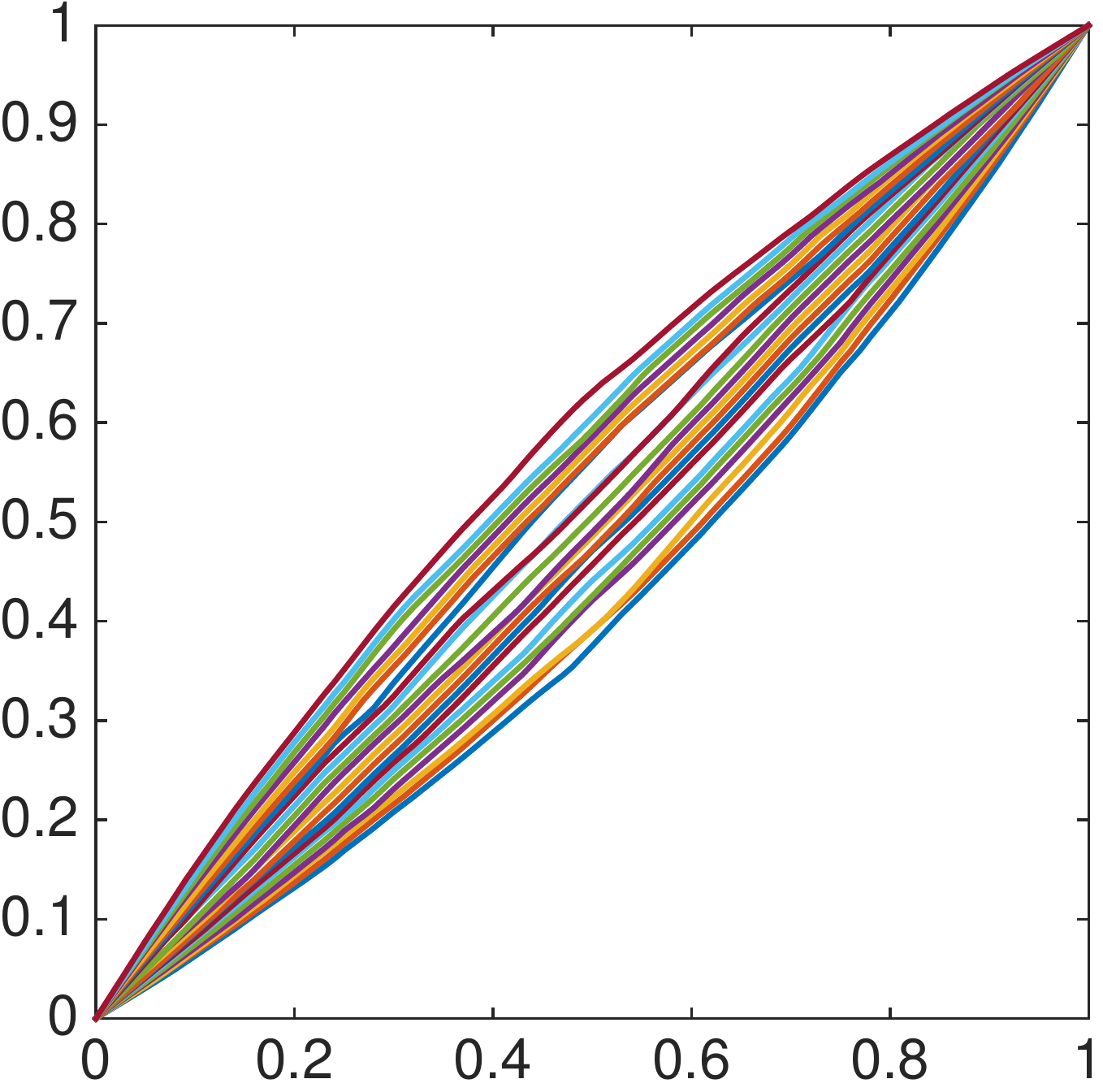} \\
      Original Functions & Amplitude & Phase 
     \end{tabular}
    \caption{Phase-amplitude separation of uncensored functional data using SRVF representation.}
    \label{fig:uncensored-align}
\end{figure}

For alignment of multiple functions, one exploits the fact that the minimum in Eqn.~\ref{eqn:elastic-dist} is actually a proper distance on the quotient space $\ltwo/\Gamma_T$. Using this distance, one can define a mean function under this metric and then align the given functions to this mean using Eqn.~\ref{eqn:elastic-dist}. For a given set of functions $\{f_1, f_2, \dots, f_n\}$, this framework results in a set of time-warping functions (also called phases) $\{ \gamma_i\}$ and the corresponding aligned functions $\{ f_i \circ \gamma_i\}$ (also called amplitudes). The full procedure for this Phase-Amplitude separation has been outlined in Chapter 8 of the textbook~\cite{srivastava-klassen:2016}. Figure~\ref{fig:uncensored-align} shows an example of this approach -- the left panel shows the original functions $\{f_i\}$, the middle panel shows the time-warped aligned functions $\{ f_i \circ \gamma_i\}$, and the right panels shows the optimal warping functions $\{ \gamma_i\}$. 

This approach has been remarkably successful in the alignment of functional data in a variety of applications. However, as mentioned earlier, this framework assumes that both the end points $t=0$ and $t=T$ are preregistered in all observations. In Fig.~\ref{fig:uncensored-align}, the endpoints are kept fixed while the interior is time-warped. Next, we consider the problem of aligning functional data with a sliding right boundary. Note that one of the boundaries, say the left one, can be matched by translation, and that leaves only the other boundary to be matched algorithmically. 

\subsection{Joint Time-Warping and Time Scaling}

Re-define the set of functions to be $\mathcal{F}=\{f:[0,\infty)\to\real \; | \; f \; \textnormal{is absolutely continuous}\}$. Define the space $\mathcal{F}_0\subset\mathcal{F}$ as the space of all absolutely continuous, right-censored functions on the non-negative reals as 
$$
\mathcal{F}_0=\{(c,f) \; | \; c\geq 0,f\in\mathcal{F},f(t)=0 \; \textnormal{for all} \; t>c\}.
$$ 
Let $(c,q)\in\ltwo_0$ be the SRVF representation of $(c,f)$. That is, $(c,q)$ represents a function $q\in\ltwo([0,\infty),\real)$ such that $q(t)=0$ for all $t>c$. 
Further, let $\ltwo_0$ be the set of SRVF representations of all 
elements of $\mathcal{F}_0$.
Define the preshape distance between censored functions as the $\ltwo$ distance on their corresponding SRVFs.
\begin{definition}
The preshape distance on $\mathcal{F}_0$ is given by the $\ltwo$ distance
$$
d_p((c_1,f_1),(c_2,f_2))^2=\int_0^{\infty}(q_1(t)-q_2(t))^2dt.
$$
\end{definition}
Note that since the two functions are censored at some point, the above integral will always be finite. Also, since the preshape distance is equal to the $\ltwo$ distance on the space $\ltwo([0,\infty),\real)$, it is indeed a proper distance.

Now, we re-define the full diffeomorphism space $\Gamma$ to be on the entire non-negative reals as such: 
$$
\Gamma=\{\gamma:[0,\infty)\to[0,\infty) \; | \; \gamma(0)=0,
$$
$$
\gamma \; \textnormal{abs. cont. and invertible}\}.
$$
It is easy to show that $\Gamma$ is a group under the binary operation of function composition and that $\Gamma$ forms a right group action on the function space $\mathcal{F}_0$ also by function composition with $(c,f)\circ\gamma=(\gamma^{-1}(c),f\circ\gamma)$. Using the definition of SRVF, one can show that the corresponding right group action of $\Gamma$ on $\ltwo_0$ is given by $(c,q)\ast\gamma = (\gamma^{-1}(c),(q\circ \gamma)\sqrt{\dot{\gamma}})$. Furthermore, similar to previous work in elastic shape analysis of functions on a fixed and finite time interval, the group $\Gamma$ acts by isometries with respect to the preshape distance $d_p$. That is, for any $\gamma\in\Gamma$ and $(c_1,f_1),(c_2,f_2)\in\mathcal{F}_0$, the distance preserving property of $d_p((c_1,f_1),(c_2,f_2))=d_p((c_1,f_1)\circ\gamma,(c_2,f_2)\circ\gamma)$ holds.

At this stage, one can use the past formulation to register functions according to $\inf_{\gamma\in\Gamma}d_p((c_1,f_1),(c_2,f_2)\circ\gamma)$. However, since our functions are zero on the right sides, the use of full $\Gamma$ is excessive. Instead, we will seek a proper subgroup of $\Gamma$ that makes the optimization more efficient and the result more meaningful. 
We construct this new group $G\subset\Gamma$ from a combination of the time-scaling and time-warping sets $H,N\subset\Gamma$, which have the following definitions.
\begin{definition}
The \textbf{time-scaling} set $H$ is given by 
$$
H=\{h\in\Gamma \; | \; h(t)=at \; \textnormal{for any} \; a>0\},
$$
the set of all linear functions on the non-negative reals. 
\end{definition}
\begin{definition}
The \textbf{time-warping} set $N$ is given by 
$$
N=\cup_{b>0}\{n\in\Gamma \; | \; n(t)=t \; \textnormal{for} \; t\geq b\},
$$
the set of fixed-interval diffeomorphisms on $[0,b]$ completed with identity from $b$ to infinity, unioned over all $b>0$. The point $b$ is termed the pivot point of the diffeomorphism $n$.
\end{definition}
$H$ and $N$ are both subgroups of $\Gamma$, with $N$ being a {\it normal} subgroup, as established next. 
\begin{lemma}
The time-scaling set $H$ is a subgroup of $\Gamma$.
\end{lemma}
\textbf{Proof}: $H\subset \Gamma$ by definition. The closure property is achieved due to linearity: $h_1\circ h_2=h_1(a_2t)=a_2h_1(t)=a_2a_1t\in H$. Any $h\in H$ has an inverse $h^{-1}(t)=t/a$, and thus $h^{-1}\in H$. The identity element $e(t)=t$ is also in $H$. $\square$
\begin{lemma}
The time-warping set $N$ is a normal subgroup of $\Gamma$.
\end{lemma}
\textbf{Proof}: $N\subset \Gamma$ by definition. To establish the closure property, note that for any $n_1,n_2\in N$, the composition $\tilde{n}=n_1\circ n_2$ is also a member of $N$, with $\tilde{b}=\max(b_1,b_2)$ such that $\tilde{n}(t)=t$ for $t\geq \tilde{b}$. 

To prove that $N$ is a {\it normal} subgroup of $\Gamma$, we need to show that $\gamma \circ n \circ \gamma^{-1} \in N$ for all $\gamma \in \Gamma$ and $n \in N$.
Evaluating the composition, we get 
$$
\gamma(n(\gamma^{-1}(t)))=\left\{\begin{array}{ll}
    \gamma(n(\gamma^{-1}(t))) & 0\leq t\leq \gamma(a) \\
    \gamma(\gamma^{-1}(t)) & t>\gamma(a)
\end{array}\right.
$$
$$
=\left\{\begin{array}{ll}
    (\gamma\circ n\circ \gamma^{-1})(t) & 0\leq t\leq \gamma(a) \\
    t & t>\gamma(a),
\end{array}\right.
$$
which is a member of $N$. $\square$

We need to form a product of $N$ and $H$ to reach the desired joint time-warping and time-scaling subgroup. However, since $N$ and $H$ do not commute, we cannot simply construct $G$ as the direct product group $G=N\times H$. Instead, we form a semidirect product. 
Using the fact that $N$ and $H$ are both groups, and that $N\cap H=\{e\}$, the identity element, we can construct a new subgroup $G\subset \Gamma$ as an {\it outer semidirect product} of both $N$ and $H$.  As a first step, we define the following homomorphism. 
\begin{definition}
Let $\phi:H\to Aut(N)$ be the group homomorphism defined by $\phi(h)(n)=h\circ n \circ h^{-1}$ for all $h\in H$ and $n\in N$. We write this homomorphism as $\phi_h:N\to N$ for brevity. 
\end{definition} 
Before we proceed, we establish that this mapping is indeed a homomorphism. 
\begin{lemma}
The map $h\mapsto\phi_h$ is a homomorphism $\phi:H\to Aut(N)$, {\it i.e.} $\phi_{h_1} \circ \phi_{h_2}=\phi_{h_1 \circ h_2}$
\end{lemma}
\textbf{Proof}: For notational brevity in this proof, we suppress the symbol $\circ$ for function composition. Since $N$ is a normal subgroup of $\Gamma$, and $H\subset \Gamma$, it holds that $\phi_{h_1}\phi_{h_2}$ and $\phi_{h_1h_2}$ are both functions from $N\to N$. For any $n\in N$, we have that 
$\phi_{h_1}\phi_{h_2}(n)=\phi_{h_1}(h_2nh_2^{-1})=h_1h_2nh_2^{-1}h_1^{-1}=(h_1h_2)n(h_1h_2)^{-1}=\phi_{h_1h_2}(n)$. $\square$

Now we have all the pieces to form the subgroup required for registration of functions. 
\begin{definition}
Using the time-warping subgroup $N$, the time-scaling subgroup $H$, and the homomorphism $\phi:H\to Aut(N)$, define the outer semidirect product $G=N\rtimes_{\phi} H$, denoted by $G\subset\Gamma$, to be the pairing $(N,H)$ with the following properties:
\begin{enumerate}
    \item Group operation: $(n_1,h_1)\cdot(n_2,h_2)=(n_1\circ\phi_{h_1}(n_2),h_1\circ h_2)$
    \item Inverse element: $(n,h)^{-1}=(\phi_{h^{-1}}(n^{-1}),h^{-1})$.
\end{enumerate} 
\end{definition}
Using Lemmas 1, 2, and 3 and the definition of the homomorphism $\phi$, by definition of an outer semidirect product, we can show that $G$ is a proper subgroup of $\Gamma$.

\textit{Remarks:} By definition, any element in $G$ can be expressed uniquely as $g=n\circ h$. (Therefore, $G$ also contains functions $g\in\Gamma$ that take the form $g(t)=at$ for all $t>b$.) 
Henceforth, we will use this shorthand notation of $g$ to indicate the time-warping and time-scaling pair $(n,h)$.

Finally, we construct the shape distance via the following optimization over the smaller group $G$ instead of the full space $\Gamma$:
\begin{definition}
The shape distance on $\mathcal{F}_0$ is given by
$$
d_s((c_1,f_1),(c_2,f_2)) = \inf_{g\in G}d_p((c_1,f_1),(c_2,f_2)\circ g)
$$
\begin{equation}
= \inf_{(n,h)\in G}d_p((c_1,f_1),(c_2,f_2)\circ (n,h)).
\label{eqn:shapedist}
\end{equation}
\end{definition}

\begin{figure}[ht]
    \centering
    \includegraphics[height=1.8in]{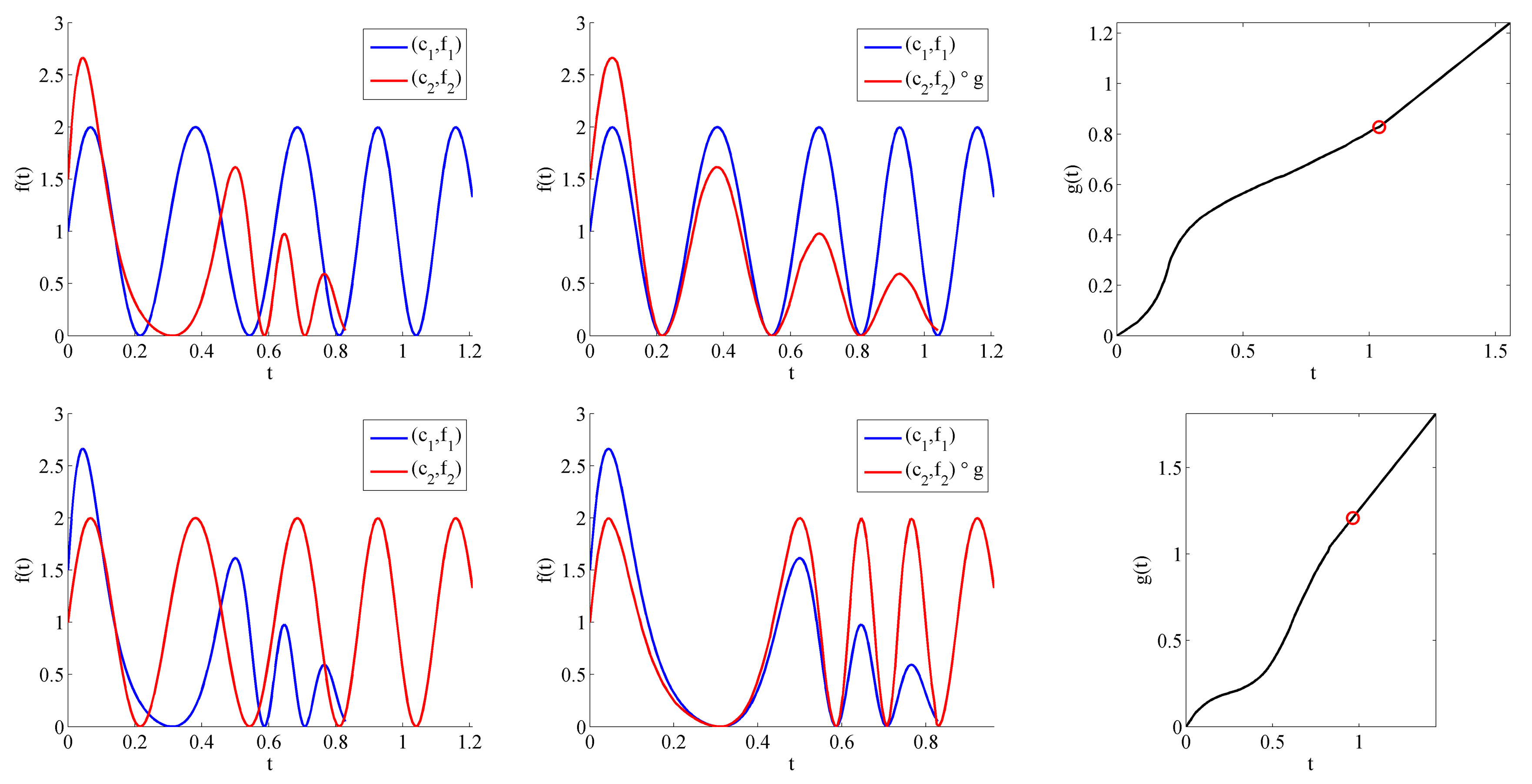}
    \caption{Alignment of two functions using elastic partial matching. Each row shows an example with the same two functions except that the labels $f_1$ and $f_2$ have been reversed in the second case. The first column shows the original functions, the second column shows the aligned functions, and the third column shows the optimal diffeomorphism $\hat{g}\in G$ used in the alignment. The red circle on each diffeomorphism represents its pivot point, which in this case is the minimum censoring point $b=\min\{c_1,\hat{c}_2\}$.}
    \label{fig:ex_shape_dist}
\end{figure}

We illustrate this idea using a simple example shown in Fig.\ \ref{fig:ex_shape_dist}. Here, we first compute the shape distance between a pair of censored functions $(c_1,f_1)$ and $(c_2,f_2)$ by aligning $(c_2,f_2)$ to $(c_1,f_1)$ according to Eq.\ \ref{eqn:shapedist}. In the first row, the left plot shows the original function pair, the center plot shows the optimally aligned $f_2$ with respect to $f_1$, and the right plot shows the optimal $g\in G$ needed to perform the time-warping of $f_2$. The second row shows the same progression of events but for when the index labeling of the two functions has been reversed. Here, since the shape distance is a proper distance and is hence symmetric, the shape distance is the same in both cases, and the optimal diffeomorphisms in $G$ are inverses of each other.

\section{Optimization Details for Pairwise Shape Distance}

Next we develop techniques for optimally aligning censored functions $(c_1,f_1)$ and $(c_2,f_2)$ in order to compute their shape distance as defined in Eq.~\ref{eqn:shapedist}. That is, we develop numerical recipes to find and apply a group element $(\hat{n},\hat{h})\in G$ to $(c_2,f_2)$ in order to optimally match a fixed $(c_1,f_1)$, minimizing the squared preshape distance. That is, we define the energy function $E$ as 
\begin{equation}
    E(n,h \,| \, (c_1,f_1),(c_2,f_2)) = d_p((c_1,f_1),(c_2,f_2)\circ(n,h))^2, 
    \label{eqn:energy_function}
\end{equation}
solve for
\begin{equation}
    (\hat{n},\hat{h})=\underset{(n,h)\in G}{\arginf} \, E(n,h) \,| \, (c_1,f_1),(c_2,f_2)),
\end{equation}
and apply $(\hat{c}_2,\hat{f}_2)=(c_2,f_2)\circ(\hat{n},\hat{h})$.

\subsection{Grid Search Algorithm} 

As a simpler first idea, we present a grid search approach for this optimization in the following manner. We start by defining a finite sampling, or ``grid,'' on the time-scaling space $H$ given by $\{h_i(t)=a_it,i=1,\ldots,J\}$. Then, for each grid point $i=1,\ldots,J$, we fix the time-scaling group element as $h_i$ and solve for the optimal time-warping $n_i$ with the pivot $b_i=\min(c_1,c_2/a_i)$. This value is the smaller of the two censoring points after the second function is time-scaled, which indicates the boundary of the common observation domain for both censored functions. One can solve for $n_i$ by first truncating the two functions to the common interval $[0,b_i]$ and then performing dense elastic registration over this interval as described in Subsection 2.1. The time-warping function $n_i$ will be completed on the right with identity outside of this interval. One can perform this function registration using one of several approaches present in the literature. The most commonly used tool is the dynamic programming algorithm~\cite{srivastava-klassen:2016}, but one can also exploit the geometry of the space $\Gamma_{b_i}$ to develop a BFGS-based gradient search~\cite{huang-BFGS-elastic}. The pair $(n_i,h_i)$ that yields the lowest value of the energy function $E$ in Eqn.\ \ref{eqn:energy_function} defines the optimal diffeomorphism $\hat{g}\in G$ that best matches $(c_2,f_2)$ to $(c_1,f_1)$.

Algorithm \ref{alg:grid_search} describes the grid search algorithm for elastic partial matching of right-censored functions. The advantages of this algorithm are that: (1) It is relatively straightforward and uses mostly existing tools from the literature, and (2) It provides global solutions depending on the grid resolution.  The disadvantage is that it is computationally expensive to solve for an optimal time-warping $n_i$ for each candidate time-scaling $h_i$. A more computationally efficient solution is to perform a gradient search over the joint domains.

\IncMargin{1em}
\SetKwInput{KwGiven}{Given}
\begin{algorithm}[ht]
\KwGiven{Censored function pair $(c_1,f_1),(c_2,f_2)$, and time-scaling grid samples $\{h_i=a_it,i=1,\ldots,J\}\in H$.}
\KwResult{Optimally aligned censored functions with associated group element and shape distance.}
Compute SRVFs $(c_1,q_1),(c_2,q_2)\in\ltwo_0$\;
\For{$i=1$ \KwTo $J$}{
    Let $b_i=\min\{c_1,c_2/a_i\}$\;
    Let $q_1'=q_11_{[0,b_i]}$ and $q_2'=(q_2\circ h_i)1_{[0,b_i]}$\;
    Solve $\hat{\gamma}_i=\underset{\gamma\in \Gamma_{b_i}}{\argmin}\|q_1'-(q_2'\circ\gamma)\sqrt{\dot{\gamma}}\|^2$ via Dynamic Programming, gradient descent, or BFGS\;
    Form $n_i$ by completing $\hat{\gamma}_i$ with identity via $n_i=\left\{\begin{array}{cc}
        \hat{\gamma}_i(t) & t\in[0,b_i] \\
        t & t>b_i
    \end{array}\right.$\;
    Let $E_i=d_p((c_1,q_1),(c_2,q_2)\circ(n_i,h_i))^2$\;
}
Let $k=\argmin_i \{E_i,i=1,\ldots,N_a\}$\;
\Return{$\hat{g}=(n_k,h_k)$, $(\hat{c}_2,\hat{f}_2)=(c_2,f_2)\circ\hat{g}$, and $d_s=\sqrt{E_k}$}
\caption{Shape Distance Via Grid Search Over $H$}
\label{alg:grid_search}
\end{algorithm}


\subsection{Gradient-Based Joint Optimization}

In order to derive a gradient-based optimization of $E$, we change to a more convenient mathematical representation for elements of the group $G$. First, notice that for $h\in H$, the mapping $\phi_h$ is an automorphism from $N$ to itself that transforms any $n\in N$ by simply scaling along the diagonal. Therefore, one can use this mapping to change the location of the pivot value $b$ without changing the overall shape of $n$. We can thus write any $n\in N$ as $n=h\circ n_0\circ h^{-1}$, where $n_0\in N$ has a pivot of $b=1$. That is, if $h(t)=bt$, then $n(t)$ can be written as $n(t)=bn_0(t/b)$, and the pivot value of $n(t)$ is $b$. By letting $n_0(t)=\gamma(t)$ on $0\leq t \leq 1$ for $\gamma\in\Gamma_1$, we can identify any $g\in G$ with the triplet $(a,b,\gamma)$ with
$$
g(t)=\left\{\begin{array}{ll}
    ab\gamma(t/b) & t\leq b \\
    at & t>b. 
\end{array}\right.
$$
However, as we have done in our implementation of the grid search algorithm, when registering two censored functions, we can simplify this parameter space a bit by setting the pivot value to $b=\min(c_1,c_2/a)$. And now since we have written $g$ in terms of two parameters $a$, and $\gamma$, we can re-write the energy function in Eqn.\ \ref{eqn:energy_function} as $E(a,\gamma \; | \; (c_1,q_1),(c_2,q_2))\triangleq E(a,\gamma)$, where 
\begin{align*}
    E(a,\gamma) &= \int_0^b(q_1(t)-q_2(ab\gamma(t/b))\sqrt{a\dot{\gamma}(t/b)})^2dt \\
    &+ \int_b^{\infty}(q_1(t)-q_2(at)\sqrt{a})^2dt.
\end{align*}
The gradient of $E$ can then be defined for both of these variables.  \\

\noindent {\bf New Mathematical Representation}:
 We will use a map $M$, given by $M(g) = (\log(a),\log(b),\sqrt{\dot{\gamma}})$, the log-transformations of $a$ and $b$ and the SRVF of $\gamma$, to reach a more convenient representation of the group $G$ for the purpose of gradient-based optimization. Let $\xi\in\real$ be such that $a=e^{\xi}$, let $\theta\in\real$ be such that $b=e^{\theta}$, and let the space of SRVFs of $\Gamma_1$ be $\Psi$. It is easy to see that for any $\gamma\in\Gamma_1$, its SRVF $\psi=\sqrt{\dot{\gamma}}$ is non-negative and has unit $\mathbb{L}^2$ norm (please refer to \cite{srivastava-klassen:2016}).
\begin{definition}
Define the space of SRVFs of $\Gamma_1$ as the positive orthant of the unit Hilbert sphere
$$
\Psi=\{\psi\in\mathbb{L}^2([0,1],\real)\;|\;\|\psi\|=1,\psi>0\;\textnormal{\it a.e.}\}.
$$
\end{definition}
Define the parameter space $\mathcal{P}=\real^2\times\Psi$ as the set of all the transformed variables $(\xi,\theta,\psi)$. Now, we can express the group $G$ as the inverse mapping $G=M^{-1}(\mathcal{P})$, whereby any $g=M^{-1}(\xi,\theta,\psi)$.

Note that ${\cal P}$ is a Lie group, with the group operation being $(\xi_1,\theta_1,\psi_1)\cdot(\xi_2,\theta_2,\psi_2)=(\xi_1+\xi_2,\theta_1+\theta_2,(\psi_1\circ \int_0^t\psi_2(s)^2ds)\psi_2)$. For any $(\xi,\theta,\psi)\in\mathcal{P}$, its inverse is given by $(-\xi,-\theta,1/\psi)\in\mathcal{P}$, and the identity element of $\mathcal{P}$ is given by $(0,0,\psi_{id})\in\mathcal{P}$ with $\psi_{id}(t)=1$ for all $t\in[0,1]$. The group action of $G=M^{-1}(\mathcal{P})$ on the censored SRVF space $\ltwo_0$ can be written in terms of the parameters $(\xi,\theta,\psi)$ as such:
$$
(c,q)\ast M^{-1}(\xi,\theta,\psi) 
$$
$$
=\left\{\begin{array}{ll}
    q(e^{\xi+\theta}(\int_0^{te^{-\theta}}\psi(s)^2ds))e^{\xi/2}\psi(te^{-\theta}) & t\leq e^{\theta} \\
    q(e^{\xi}t)e^{\xi/2} & t>e^{\theta},
\end{array}\right. 
$$
with $c\mapsto (M^{-1}(\xi,\theta,\psi))^{-1}(c)$. Now, the energy function can be re-written as $E(\xi,\psi \; | \; (c_1,q_1),(c_2,q_2))\triangleq E(\xi,\psi)$ with
\begin{align*}
    E(\xi,\psi) &= d_p((c_1,q_1),(c_2,q_2)\ast M^{-1}(\xi,\log(b),\psi))^2 \\
    &= \int_0^b(q_1(t)-q_2(e^{\xi}b(\int_0^{t/b}\psi(s)^2ds))e^{\xi/2}\psi(t/b))^2dt \\
    &+ \int_b^{\infty}(q_1(t)-q_2(e^{\xi}t)e^{\xi/2})^2dt,
\end{align*}
where $b=\min(c_1,c_2e^{-\xi})$. Note that here we have set the variable $\theta$ to be a function of $\xi$ and $c_1,c_2$, and thus it does not appear in the argument of $E$. Also, this implies that with respect to the group action $\ast$, the censoring point $c_2$ changes to $c_2e^{-\xi}$. 

The reason for changing the representation of transformations from $G$ to ${\cal P}$ is that the Riemannian geometry of $\mathcal{P}$ is less complex, relatively, in that we know the expressions for tangent spaces, exponential maps and geodesics in ${\cal P}$. An additional nice property of $\mathcal{P}$ that we will exploit later in our optimization algorithm is that it is a Lie group. The Lie group structure allows us to formulate update vectors at each iteration in the tangent space of the identity element and apply these updates to the current estimate in a sequential manner. For now, since in our implementation we are setting $\theta$ as a function of $\xi$, we remove this parameter from the explicit representation of an element in $\mathcal{P}$. 
\begin{definition}
Define the tangent space of $\mathcal{P}$ at the identity element $p_{id}=(0,1)$ as
$$
T_{p_{id}}(\mathcal{P})=\left\{v=(y,z)\in\real\times\ltwo([0,1],\real)\;|\;\int_0^1z(t)=0\right\}.
$$
\end{definition}
\begin{definition}
Define the exponential mapping of $v=(y,z)\in T_{p_{id}}(\mathcal{P})$ as $\exp_{p_{id}}(v)\in\mathcal{P}$ via the formula
$$
\exp_{p_{id}}(v)=(y,\cos(\|z\|)+\sin(\|z\|)\frac{z}{\|z\|}).
$$
\end{definition}

In an iterative line-search optimization method, for the iteration $k$, let the current estimate be $p_k=(\xi_k,\psi_k)\in\mathcal{P}$, a search direction vector $v_k$ be in the tangent bundle of $\mathcal{P}$, and a step size be $\alpha_k>0$. If $\mathcal{P}$ were a vector space, the updated estimate would be given simply by $p_{k+1}=p_k+\alpha_kv_k$. Since $\mathcal{P}$ is instead a non-linear manifold, one can compute the update using the exponential mapping $p_{k+1}=\exp_{p_k}(\alpha_kv_k)$. However, since $\mathcal{P}$ is also a Lie group with an identity element and a group action, there is no need to compute search directions in any tangent space except at identity. For any $k$, we can compute $v_k\in T_{p_{id}}(\mathcal{P})$ and then apply a sequential update to $p_k$ via $p_{k+1}=p_k\cdot\exp_{p_{id}}(\alpha_kv_k)$, where $\cdot$ is the group operation of $\mathcal{P}$. 

Most often, the search direction $v_k$ is based on the gradient of the objective function, or energy function, evaluated at the current estimate. Therefore, we need to define the energy $E_k:\mathcal{P}\to\real$ at step $k$ and show how to compute its gradient as an element of $T_{p_{id}}(\mathcal{P})$. First, define the current censored function as $(c_{2,k},q_{2,k})\triangleq(c_2,q_2)\ast M^{-1}(\xi_k,\log(b_k),\psi_k)$, where $b_k=\min(c_1,c_2e^{-\xi_k})$. 
\begin{definition} \label{def:E}
For an update $(\xi,\psi)\in\mathcal{P}$ at iteration $k$, define the energy of updating the current censored function $(c_{2,k},q_{2,k})$ by $(\xi,\psi)$ as $E_k(\xi,\psi)$
\begin{eqnarray*}
     &=& d_p((c_1, q_1), (c_{2,k},q_{2,k}) \ast M^{-1}(\xi,\log(b),\psi))^2\\
    &=&\int_0^b(q_1(t)-q_{2,k}(e^{\xi}b(\int_0^{t/b}\psi(s)^2ds))e^{\xi/2}\psi(t/b))^2dt \\
    &+& \int_b^\infty \left(q_1(t)-q_{2,k}(e^{\xi}t)e^{\xi/2}\right)^2dt,
\end{eqnarray*}
where $b=\min(c_1,c_{2,k}e^{-\xi})$.
\end{definition}

\noindent {\bf Derivation of Gradient of $E$}: \\
Now, we compute the gradient $\nabla E_k$ at identity and define a line-search update direction vector $v_k\in T_{p_{id}}(\mathcal{P})$ based on this gradient. Before writing the full analytical expression for the gradient, we develop a series of useful results. 
\begin{lemma} \label{lemma1}
For $(c_1,q_1),(c_2,q_2)\in\ltwo_0$ and $(\xi,\log(b),\psi)\in\mathcal{P}$ with $b=\min(c_1,c_2e^{-\xi})$, let $\tilde{q}_2\triangleq(c_2,q_2)\ast M^{-1}(\xi,\log(b),\psi)$. The derivative of $\tilde{q}_2$ with respect to $\xi$ evaluated at the identity element $(\xi_{id},\psi_{id})\in\mathcal{P}$ is given by
$\frac{\partial\tilde{q}_2}{\partial\xi}(\xi_{id},\psi_{id})=t\dot{q}_2+\frac{1}{2}q_2$ for all $t\geq 0$.
\end{lemma}
\textbf{Proof}: See Appendix A. 
\begin{lemma} \label{lemma3}
Let $(c_1,q_1),(c_2,q_2)\in\ltwo_0$, let $b=\min(c_1,c_2)$ and let $F(\psi)=\int_0^b(q_1(t)-(q_2(b\int_0^{t/b} \psi(s)^2 ds) \psi(t/b))^2dt$. Then, if $x=t/b$, the Riemannian gradient at identity $\nabla F\in T_{\psi_{id}}(\Psi)$ is given by
$
\nabla F = w(x)-\int_0^1w(x)dx$,
with 
$w(x)=4b^2\int_0^x(q_1(bs)-q_2(bs))\dot{q}_2(bs)ds-2b(q_1(bx)-q_2(bx))q_2(bx)$.
\end{lemma}
\textbf{Proof}: See Appendix B. 

Having obtained these useful results, we now derive the expression for gradient of $E_k$ with respect to an incremental element of ${\cal P}$.
\begin{thm}
At iteration $k$ and for $(c_1,q_1),(c_{2,k},q_{2,k})\in\ltwo_0$, the gradient of $E_k$ at identity $(\xi_{id},\psi_{id})\in\mathcal{P}$ is written as the pair $\nabla E_k(\xi_{id},\psi_{id})=(\frac{\partial E_k}{\partial\xi}(\xi_{id},\psi_{id}),\frac{\partial E_k}{\partial\psi}(\xi_{id},\psi_{id})) \in T_{p_{id}}({\cal P})$. The two terms that comprise the gradient vector are defined as follows.
\begin{enumerate}
    \item The partial derivative of $E_k$ with respect to $\xi\in\real$ evaluated at identity is given by $\frac{\partial E_k}{\partial\xi}(\xi_{id},\psi_{id})$
    \begin{equation}
    =-2\int_0^{\infty}(q_1-q_{2,k})(t\dot{q}_{2,k}+\frac{1}{2}q_{2,k})dt \in \real.
    \label{eqn:theorem-eq1}
    \end{equation}
    \item Let $b=\min(c_1,c_2)$, $x=t/b$, and define the function $w_k \in T_{\psi_{id}}(\Psi)$ as
    \begin{align*}
        w_k(x) &= 4b^2\int_0^x(q_1(bs)-q_{2,k}(bs))\dot{q}_{2,k}(bs)ds \\
        &- 2b(q_1(bx)-q_{2,k}(bx))q_{2,k}(bx)\ .
    \end{align*}
    Then, the partial derivative of $E_k$ with respect to $\psi\in\Psi$ evaluated at identity is given by
    \begin{equation}
    \frac{\partial E_k}{\partial\psi}(\xi_{id},\psi_{id})=w_k(x)-\int_0^1w_k(x)dx.
    \label{eqn:theorem-eq2}
    \end{equation}
\end{enumerate}
\end{thm}
\textbf{Proof}: For part (1), define the function $\tilde{q}_{2,k}=(c_{2,k},q_{2,k})\ast M^{-1}(\xi,\log(b),\psi)$, where $b=\min(c_1,c_{2,k}e^{-\xi})$. Now, we compute the partial derivative as 
$$
\frac{\partial E_k}{\partial\xi} = -2\int_0^{\infty}(q_1-\tilde{q}_{2,k})\frac{\partial\tilde{q}_{2,k}}{\partial\xi}dt. 
$$
Using Lemma 4, the above expression evaluated at identity becomes the expression given in Eqn.~\ref{eqn:theorem-eq1}.

For part (2), define $F_k(\psi)=E_k(\xi_{id},\psi)-\int_b^{\infty}(q_1(t)-q_{2,k}(t))^2dt$. Notice that $F_k(\psi)$ takes the form of $F$ as defined in Lemma 5 with $q_2$ replaced with $q_{2,k}$, and by construction $\nabla F_k(\psi_{id})=\frac{\partial E_k}{\partial\psi}(\xi_{id},\psi_{id})$. Thus, according to Lemma 5, the partial derivative is equal to the expression given in Eqn.~\ref{eqn:theorem-eq2}.  $\Box$

In order to implement a backtracking line-search method based on the Armijo-Goldstein condition, we must first define an inner product on $T_{p_{id}}(\mathcal{P})$.
\begin{definition}
For $v_1,v_2\in T_{p_{id}}(\mathcal{P})$ with $v_1=(y_1,z_1)$ and $v_2=(y_2,z_2)$, the chosen inner product on $\mathcal{P}$ is given by $\langle\langle v_1,v_2 \rangle\rangle=y_1y_2+\langle z_1,z_2 \rangle$, where $\langle \cdot,\cdot \rangle$ is the standard $\ltwo$ inner product, and the corresponding norm is given by $\|v\|_{\mathcal{P}}=\sqrt{\langle\langle v,v \rangle\rangle}$.
\end{definition}
With this inner product and norm for elements of the tangent space, we can define the Armijo-Goldstein condition for the backtracking line-search method. 
\begin{definition}
\textbf{(Armijo-Goldstein Condition)} For a candidate update $p=(\xi,\psi)\in\mathcal{P}$, a scalar $\beta\in(0,1)$, a search direction $v_k\in T_{p_{id}}(\mathcal{P})$, and a stepsize $\delta>0$, the Armijo-Goldstein condition is given by $E_k(p)\leq E_k(p_{id})+\frac{\beta\delta}{\|v_k\|_{\mathcal{P}}}\langle\langle\nabla E_k(p_{id}),v_k\rangle\rangle$. 
\end{definition}
One can see that in the special case of gradient descent where $v_k=-\nabla E_k(p_{id})$, the above condition simplifies to $E_k(p)\leq E_k(p_{id})-\beta\delta\|\nabla E_k(p_{id})\|_{\mathcal{P}}$. In addition to the Armijo-Goldstein condition, we also need the condition that $(\psi_k\ast\psi)(t)\geq 0$ for all $t\in[0,1]$ in order to ensure that the incremental update of $\psi_k$ remains in $\Psi$. If either of the two conditions are not satisfied, then one must reduce the stepsize by a factor of $\tau\in(0,1)$ by updating $\delta\mapsto\tau\delta$ until both are satisfied or until $\delta$ becomes too small.

Algorithm \ref{alg:gradient_descent} outlines the gradient descent method for elastic partial matching of functions. 
\begin{algorithm}[ht]
\KwGiven{A pair of censored functions $(c_1,q_1),(c_2,q_2)\in\ltwo_0$, tolerance $\epsilon>0$, maximum number of iterations $K$, scalar parameters $\beta,\tau\in(0,1)$, and stepsize $\delta>0$.}
\KwResult{Optimally aligned censored functions.}
Initialize variables $k=0$, $p_0=p_{id}$, and $(c_{2,0},q_{2,0})=(c_2,q_2)$\;
Compute $E_0(p_{id})$ via Eq.\ \ref{eqn:energy_function} and $\nabla E_0(p_{id})$ via Eqs.\ \ref{eqn:theorem-eq1} and \ref{eqn:theorem-eq2}\;
\While{$\|\nabla E_k(p_{id})\|_{\mathcal{P}}>\epsilon$ and $k<K$}{
Let $v_k=-\nabla E_k(p_{id})$\;
Let $\delta_k=\delta$\;
Generate candidate update $p=\exp_{p_{id}}(\delta_k v_k)$\;
Compute $E_k(p)$\;
\While{$E_k(p)>E_k(p_{id})-\beta\delta_k\|\nabla E_k(p_{id})\|_{\mathcal{P}}$ or $(\psi_k\odot\psi)<0$ anywhere}{
Let $\delta_k=\tau\delta_k$\;
Generate candidate update $p=\exp_{p_{id}}(\delta_k v_k)$\;
Compute $E_k(p)$\;
}
Update $(c_{2,k+1},q_{2,k+1})=(c_{2,k},q_{2,k})\ast M^{-1}(\xi,\psi)$, and update $p_{k+1}=p_k\cdot p$\;
Update $k=k+1$\;
}
\Return{$(\hat{c}_2,\hat{q}_2)=(c_{2,k},q_{2,k})$ and $\hat{g}=M^{-1}(p_k)$.}
\caption{Gradient Descent for Alignment of Right-Censored Functions.}
\label{alg:gradient_descent}
\end{algorithm}

\subsection{Modification of Energy Function}

The energy defined above (Definition \ref{def:E}) depends in part on the $\ltwo$ norm of the extra portion of the longer function (the second term). Therefore, if that piece has a large $\ltwo$ norm, then it tends to dominate the total energy. In that case, an optimal alignment pushes the endpoints of the two functions to be closer together, and the results look more similar to the standard elastic registration with identical boundaries. In order to control the influence of this unmatched part, we can modify the energy function $E$ as follows.

Multiply the second term in $E$ by a constant $\lambda > 0$, 
resulting in: 
\begin{align}
    E(\xi,\psi) &= \int_0^b(q_1(t)-q_2(e^{\xi}b(\int_0^{t/b}\psi(s)^2ds))e^{\xi/2}\psi(t/b))^2dt \notag \\
    &+ \lambda\int_b^{\infty}(q_1(t)-q_2(e^{\xi}t)e^{\xi/2})^2dt,
    \label{eq:Emodified}
\end{align}
where $b=\min(c_1,c_2e^{-\xi})$. The resulting gradient expressions are altered in only a minor fashion and for the sake of brevity are not repeated here. 

We note that for $\lambda \neq 1$, the energy $E$ is no longer the square of a proper metric, and some of the nice mathematical properties of a quotient space metric are lost. The energy minimization process results in a measure of ``dissimilarity'' rather than a proper distance squared. One can still perform a clustering analysis with these tools, as we demonstrate in the next Section.   

\section{Experimental Results}

In this section we present some experimental results demonstrating the strengths of this framework in comparing functions with variable right boundaries. We first introduce the two functional data sets -- one simulated and one real -- that we use to demonstrate results and test our method against other existing methods. 
\\

\begin{figure}[ht]
    \centering
    \includegraphics[width=1\linewidth]{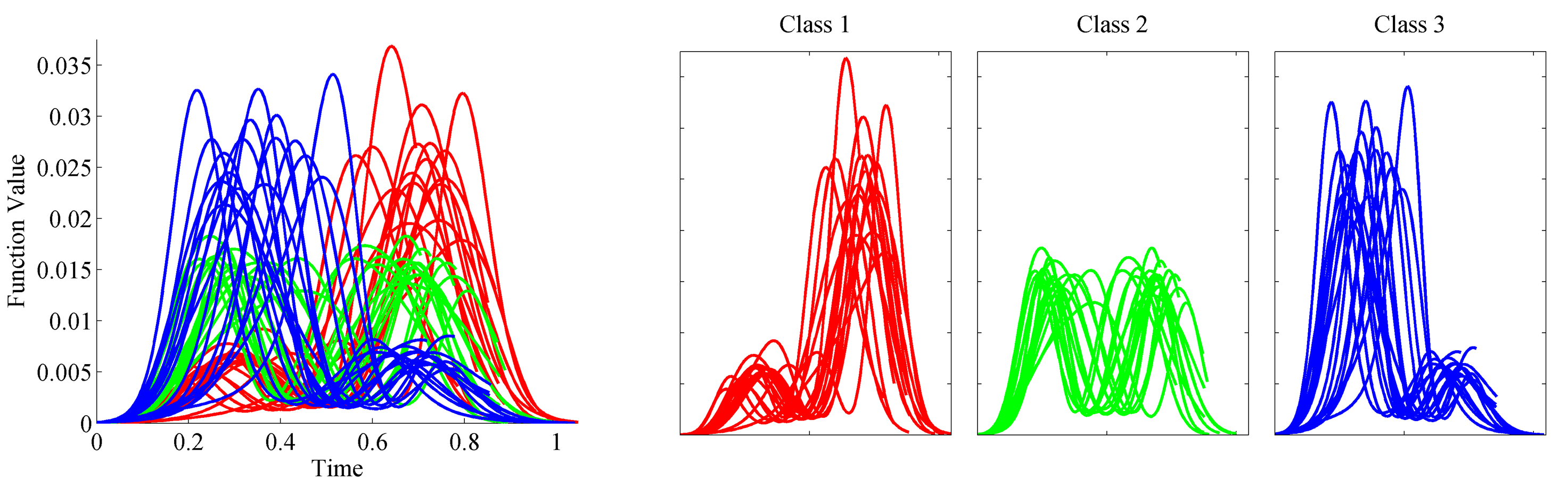}
    \caption{Simulated data set with three classes. The left-most panel shows all functions plotted in the same window and colored according to class label. The remaining three panels plot each of the three classes separately.}
    \label{fig:sim_data}
\end{figure}

\noindent{}
{\bf Dataset 1}: The simulated data set consists of $51$ functions, each with $156$ sample points, and separates into three classes of $17$ functions each. The functions are all based on a mixture of two Gaussians on the interval $[0,1]$ with means at $t=0.3$ and $t=0.7$ and with the same variance. The functions within each class have the same mixture coefficients, where, for class 1, 2 and 3, we use the coefficient pairs $(0.2,0.8)$, $(0.5,0.5)$, and $(0.8,0.2)$, respectively. To generate a function $f$ in the simulated dataset, we start with the appropriate Gaussian mixture on $[0,1]$ and alter it in the following manner: (1) select $b\sim Uniform[0.7,1]$ and truncate $f$ to the interval $[0,b]$; (2) select $a\sim Uniform[0.9,1.1]$ and apply the time-scaling function $h(t)=at$ to $f$; and (3) apply a random time-warping diffeomorphism in $\Gamma_{b/a}$ to $f$. Fig.\ \ref{fig:sim_data} shows the resulting simulated data set, colored according to class label.   
\\

\begin{figure}[ht]
    \centering
    \includegraphics[width=\linewidth]{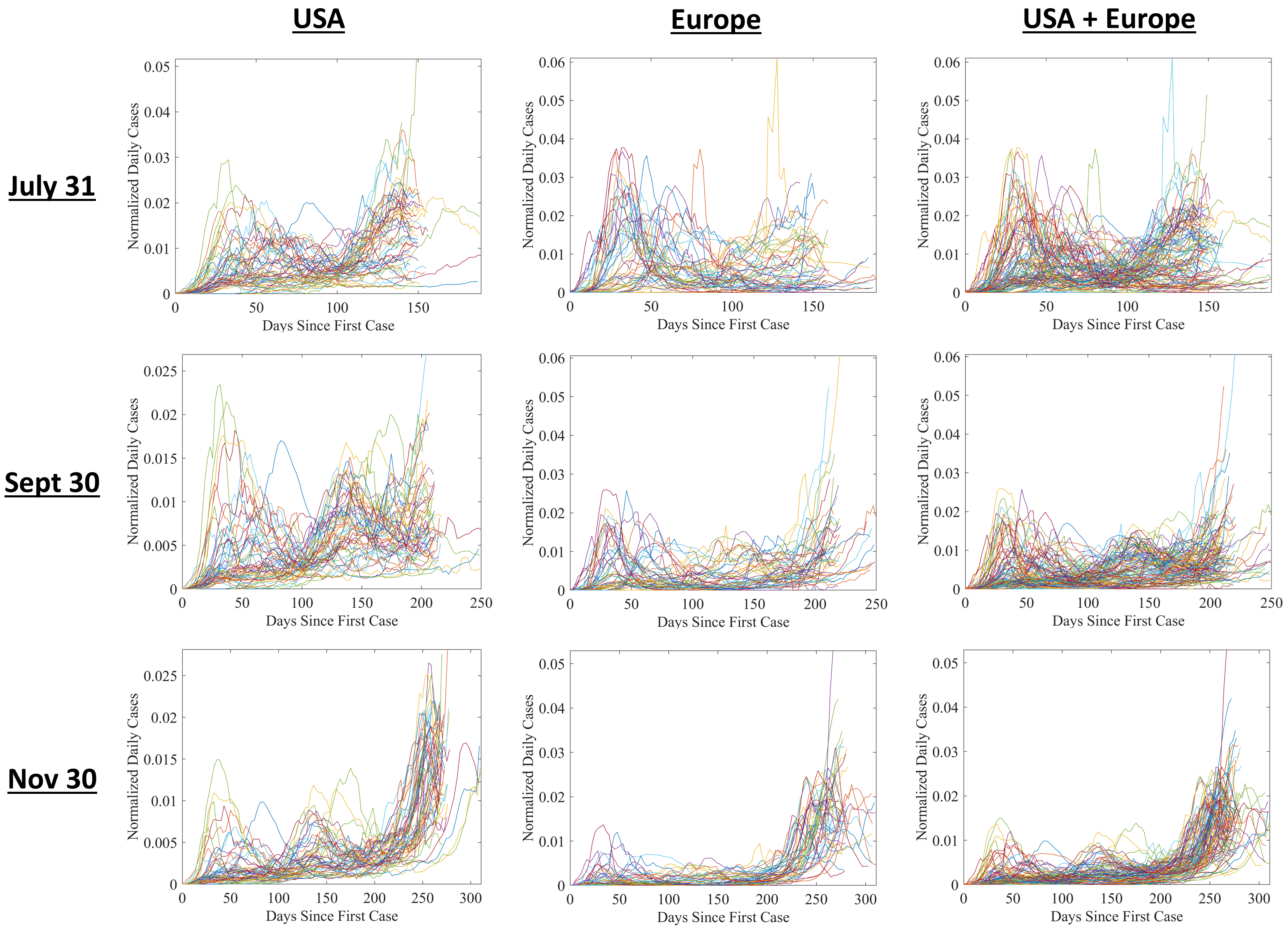}
    \caption{United States and European COVID-19 normalized infection rate curves truncated at three different right boundaries -- July 31, September 30, and November 30. For each panel, the time axis has a unit of days and indicates the number of days since the first recorded case. In the first and second columns, the US and European data is plotted separately for ease of visualization, and in the third column they are plotted together.}
    \label{fig:covid_data}
\end{figure}

\noindent
{\bf Dataset 2}: The real data used here comes from COVID-19 virus infection rate curves for each of the 50 United States plus the District of Columbia and Puerto Rico plus 47 European countries. The daily new case data is preprocessed via the following procedure. First, we assign the time domain to have a unit of days, and then we translate each curve so that $t=0$ represents the day of the first recorded case. Thus, $t\in [0,\infty)$ represents the number of days since a state or country's first case. Then, we apply a seven day moving average to smooth the data, and then we re-sample each curve via spline interpolation to have 100 uniformly spaced time sample points. Finally, we normalize the rate curve so that it has integral $1$. We execute this process three times, each on raw data truncated at different ending dates -- July 31, September 30, and November 30 -- to create three datasets of normalized infection rate curves. Fig.\ \ref{fig:covid_data} shows a plot of all $99$ normalized infection rate curves for the United States and Europe. The first two columns separate the US and European curves, respectively, for visual purposes.

\subsection{Pairwise Alignment Examples}

\begin{figure}[ht]
    \centering
    \includegraphics[width=1\linewidth]{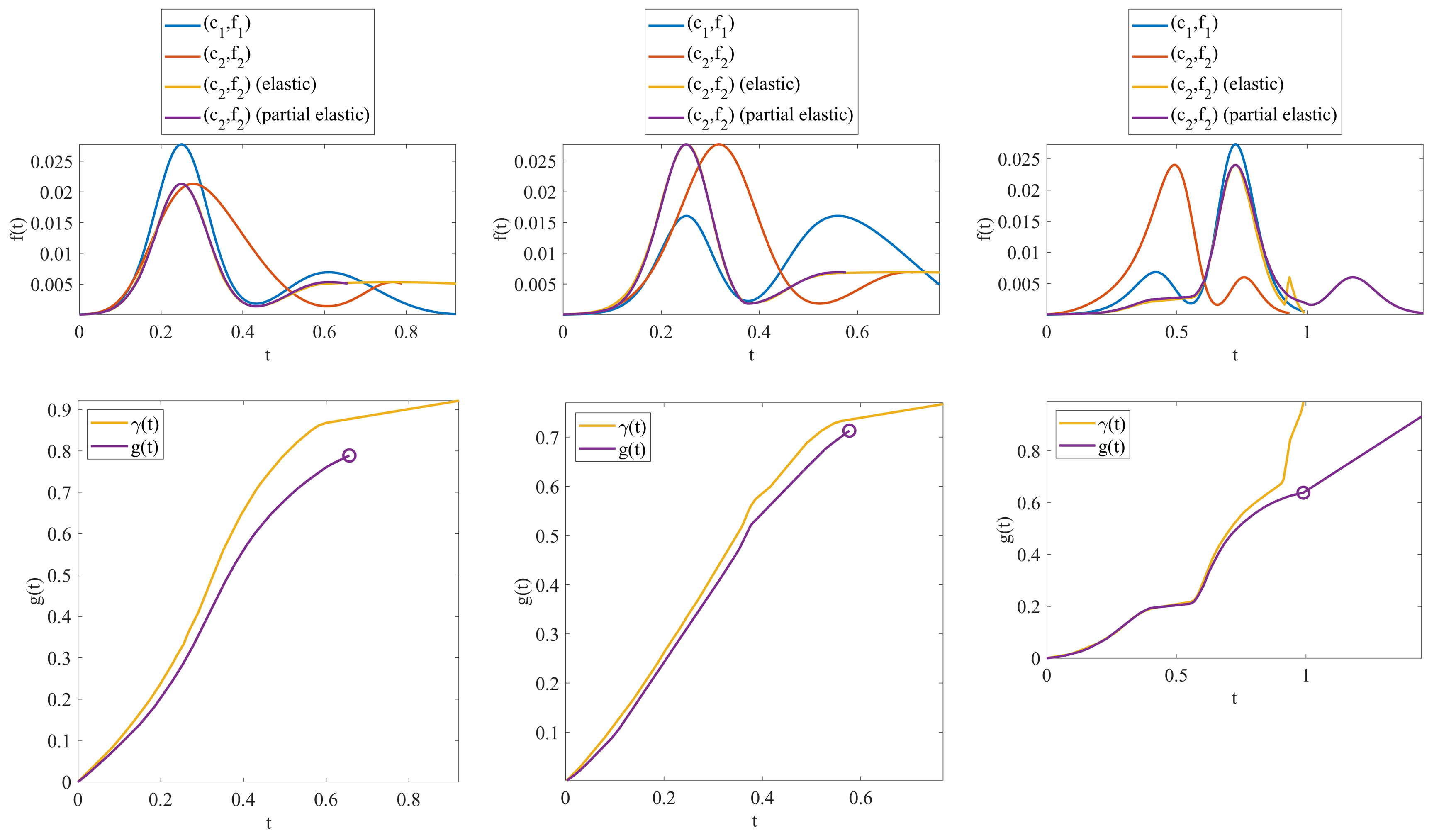}
    \includegraphics[width=1\linewidth]{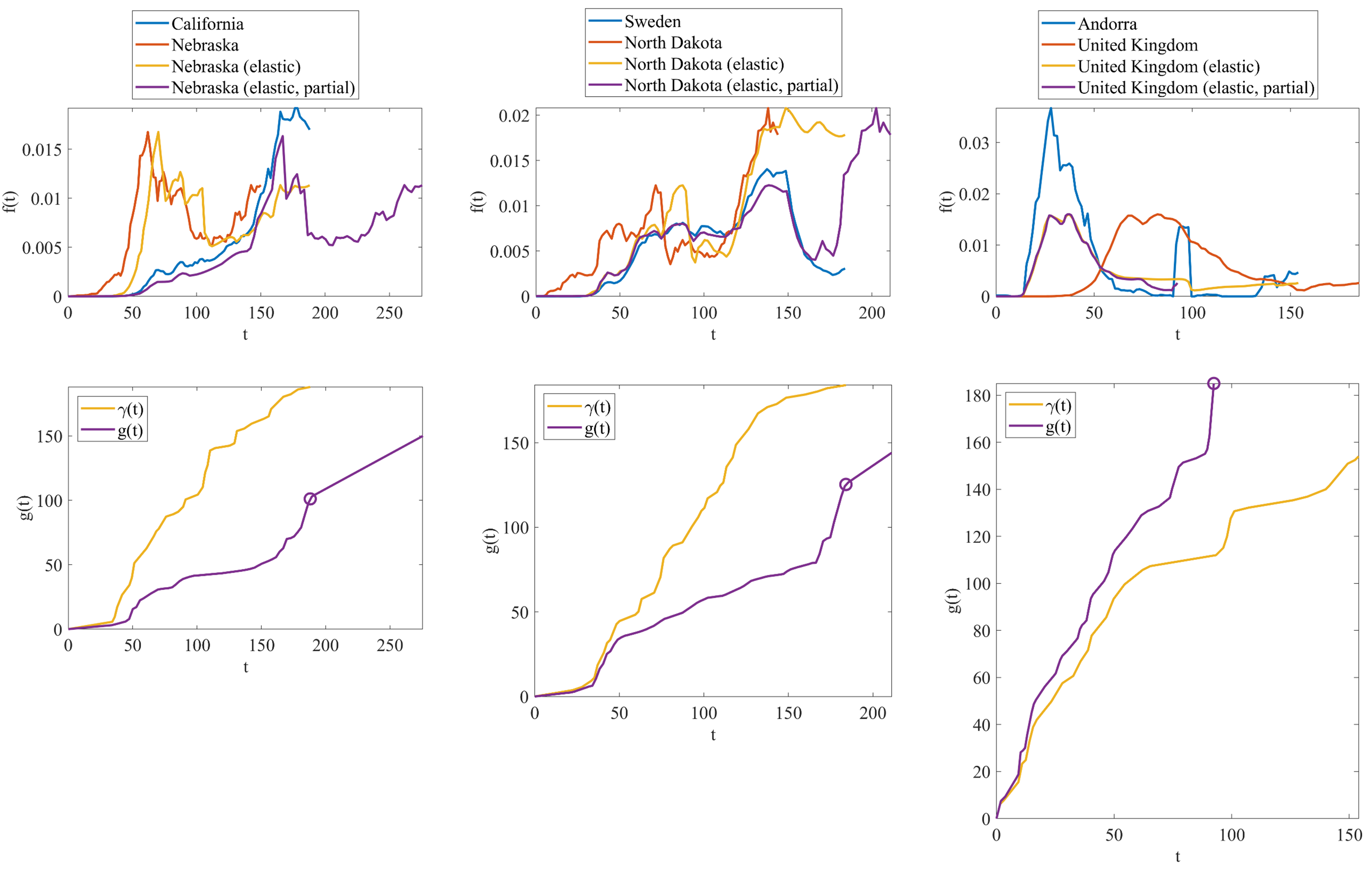}
    \caption{Examples of pairwise elastic function registrations. The first row shows three pairwise alignment examples for the simulated dataset. In each panel, the blue and red curves are the original censored functions $(c_1,f_1)$ and $(c_2,f_2)$, the yellow curve is the aligned version of $(c_2,f_2)$ using standard elastic registration with fixed and matched endpoints, and the purple curve is the aligned version of $(c_2,f_2)$ using our novel elastic partial matching methodology with floating right endpoint. The second row provides the associated optimal diffeomorphisms that achieve the alignments, plotted in the same color scheme and time axis scale. The third and fourth rows are the same as the first and second but instead using examples from the COVID-19 data set. Here, we restrict example pairs to be from the set of US states and European countries truncated at July 31.}
    \label{fig:pairwise_examples}
\end{figure}

Fig.\ \ref{fig:pairwise_examples} shows a few interesting examples of pairwise alignments for both the simulated and the real data sets. For each function pair we compute both the standard elastic alignment with fixed and matched endpoints and our novel elastic partial matching with floating endpoints. For the partial matching, we use the modified energy function in Eq.\ \ref{eq:Emodified} with $\lambda=0.25$. In each panel, we plot the original censored functions $(c_1,f_1)$ and $(c_2,f_2)$ in blue and red, respectively; the elastically registered version of $(c_2,f_2)$ in yellow with matched right endpoint; and the partially matched version via the methodology developed in Sections 2 and 3 in purple. In each case, the alignments are all noticeably different visually, with our elastic partial matching methodology providing a more natural registration than the standard elastic methodology. For some examples, we see that the curve with standard elastic registration is similar to our partially matched curve on a portion of the domain. However, as it approaches the endpoint of the fixed blue curve, the elastic registration tends to unnaturally compress or stretch the yellow curve in order to force the right endpoints to match. In other cases, the freedom offered by our partial matching methodology allows us to find a completely different matching across the entire domain. In particular, the elastic partial matching of North Dakota to Sweden's infection rate curve allows for a surprisingly similar shape matching that the standard elastic registration was unable to uncover.

Below each pairwise matching result, we also show the corresponding optimal diffeomorphisms for the two elastic methods. Using the same color scheme and the same domain (time axis) scale, we plot the optimal $\hat{\gamma}\in\Gamma_{\hat{b}}$ and $\hat{g}\in G$ associated with the two elastic registrations. The circle along the purple colored diffeomorphism $\hat{g}\in G$ represents the point $\hat{g}(\hat{b})$, where $\hat{b}=\min\{c_1,\hat{c}_2\}$ is the pivot point. Recall that beyond this point the function $\hat{g}$ is linear with slope $\hat{g}(\hat{b})/\hat{b}$ and extends to infinity. In cases where the optimally matched second function (purple) is shorter than the first (blue), we do not plot $\hat{g}(t)$ beyond the pivot point $t=\hat{b}$.

\subsection{Bayesian Clustering of Functional Data}

Next, we perform unsupervised clustering of the data using three different pairwise distance (or dissimilarity) measures and compare the results. The three methods are : (a) the standard $\ltwo$ metric with fixed and matched right endpoints, (b) the standard elastic metric with fixed and matched right endpoints, and (c) our novel elastic partial matching of SRVFs with floating right endpoint. Note that method (a) operates on the function values themselves while methods (b) and (c) use SRVFs for comparison. For method (c), we use the grid search method (Algorithm 1) with $J=50$ logarithmically uniformly spaced time-scaling grid samples $\{a_i,i=1,\ldots,J\}=\{0.5,\ldots,2\}$ to initialize Algorithm 2, the gradient descent method for local refinement. Here, we use a convergence tolerance of $\epsilon=10^{-4}$, initial step size of $\delta=10^{-4}$, and standard backtracking parameters of $\beta=0.1$ and $\tau=0.5$. Finally, for method (c) we use a weighting coefficient of $\lambda=0.25$, and thus the pairwise comparison in this method is not a proper distance but rather a more general dissimilarity measure. Methods (a) and (b) both yield proper distances; however, as described next, our chosen clustering method depends on a similarity matrix instead of a distance/dissimilarity matrix, so it is not necessary for the analysis that our method of pairwise comparison be a proper distance.  

We use the clustering technique described in \cite{zhang2015} to cluster the data for each case. This clustering method is Bayesian in nature and uses only a pairwise similarity matrix as input -- not the functional data itself -- to determine the optimal number of clusters and the cluster members in an MCMC approach. It assumes a Wishart prior on the similarity matrix and uses a variant of the Chinese Restaurant Process to help determine the number of clusters. In the associated clustering software there are several hyperparameters that must be set; however, we simply use all default settings provided in the code, with an initial number of clusters set to 3 for the simulated data and 10 for the real data, to collect our clustering results. Since the software takes as input a pairwise similarity matrix $S$ instead of a distance/dissimilarity matrix $D$, we must compute a full pairwise dissimilarity matrix for all function pairs in the dataset and convert it to a pairwise similarity matrix before executing the clustering algorithm. Here, we use the conversion formula $S_{ij}=1-D_{ij}/\max\{D\}$ so that all entries of the similarity matrix $S$ are scaled similarly between 0 and 1. 
\\

\begin{figure*}[t]
    \centering
    \includegraphics[width=\linewidth]{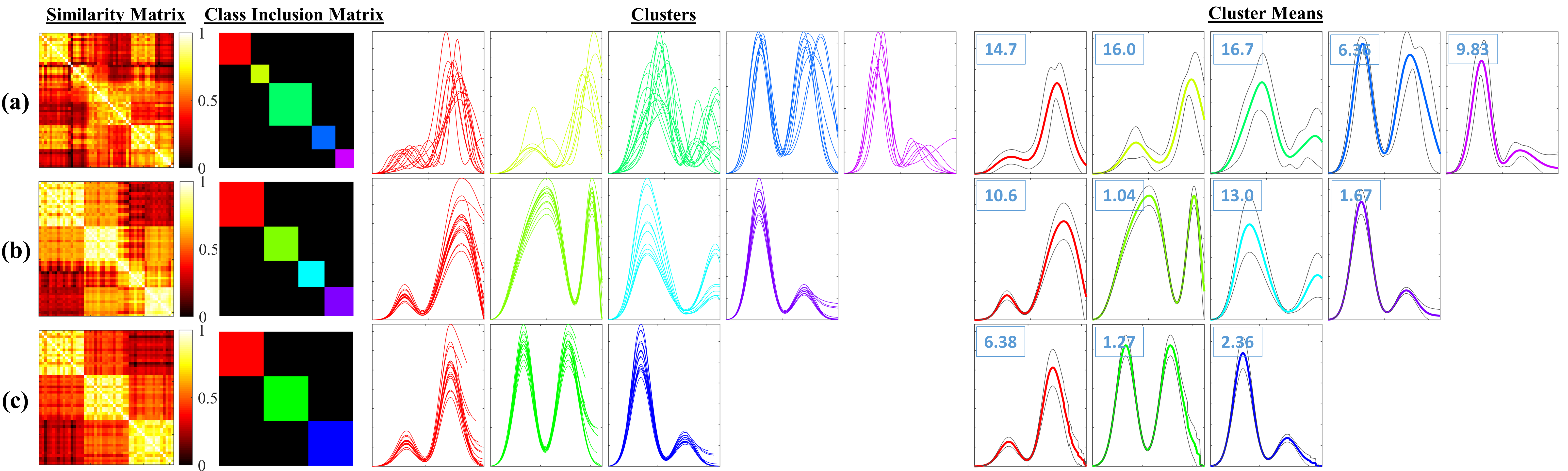}
    \caption{Simulated data set classes resulting from the implementation of the Bayesian clustering algorithm. Row (a) shows the results for the $\ltwo$ metric with fixed and identical endpoints; Row (b) shows the results for the standard elastic metric with fixed and identical endpoints; and Row (c) shows results for our novel elastic partial matching. The first two columns show the pairwise similarity matrices and the corresponding block diagonal class inclusion matrices, colorized according to class label. The center panels are the individual clusters plotted separately, with each $y$-axis scaled so that the maximum value is equal to that of the maximum $y$-value of the associated cluster. For rows (b) and (c), plots of individual classes show function members that have been mutually aligned within classes. The right-most set of panels shows the cross-sectional mean functions for each cluster with cross-sectional error bounds of $\pm$ one standard deviation. The number in the upper-left corner of each panel here represents the average cross-sectional variance for that cluster ($\times 10^{-6}$).}
    \label{fig:sim_clusters}
\end{figure*}

\noindent
{\bf Dataset 1}:
Fig.\ \ref{fig:sim_clusters} shows the result of the Bayesian clustering on the simulated dataset for each of the three methods: (a), (b), and (c). From left to right, we show the pairwise similarity matrices, the color-coded block diagonal class inclusion matrices, the function clusters with consistent color scheme, and the associated cluster means for each cluster. In order to form block diagonal matrices, the data set indices have been optimally permuted as a result of the clustering algorithm.  One can see that the number of clusters (3) is correct only for the elastic partial matching method shown in row (c). Furthermore, the cluster labeling is 100\% correct for this method. In both methods (b) and (c), the tiled cluster display shows functions that have been mutually aligned according to their respective alignment methods. For method (c), the mutual alignment is performed so that it iteratively seeks out the ``longest'' censored function and aligns each remaining function to it. For ease of visual display and the minimization of white-space, the vertical axis has been scaled in each cluster tile so that the maximum value is equal to the maximum function value in each cluster set. One can see that the functions align well within the clusters for our method, showcasing the three distinct classes. Since the data is partially observed, the other two methods tend to mislabel and/or divide the three classes unnecessarily into further subclasses.  

The right-most panels in Fig.\ \ref{fig:sim_clusters} show the cross-sectional pointwise mean for each of the clusters shown in the center. Additionally, we compute the cross-sectional variance and plot grey lines to represent plus and minus one cross-sectional standard deviation from the mean. The number in the top left of each panel window here represents the average cross-sectional variance for that cluster.
As is evident from these numbers, the standard $\ltwo$ analysis leads to artificially inflated variances due to lack of alignment. The standard elastic analysis helps to bring down the variance within clusters by aligning peaks and valleys, but since the data is partially observed, the restriction of identical right endpoints still drives up the variance unnecessarily. The average cross-sectional variances are $12.7 \times 10^{-6}$ for method (a), $6.58 \times 10^{-6}$ for method (b), and $3.34 \times 10^{-6}$ for method (c). Thus, our elastic partial matching method yields the lowest average cross-sectional variance of the three methods and the tightest clustering.
\\

\begin{figure}[ht]
    \centering
    \includegraphics[height=2in]{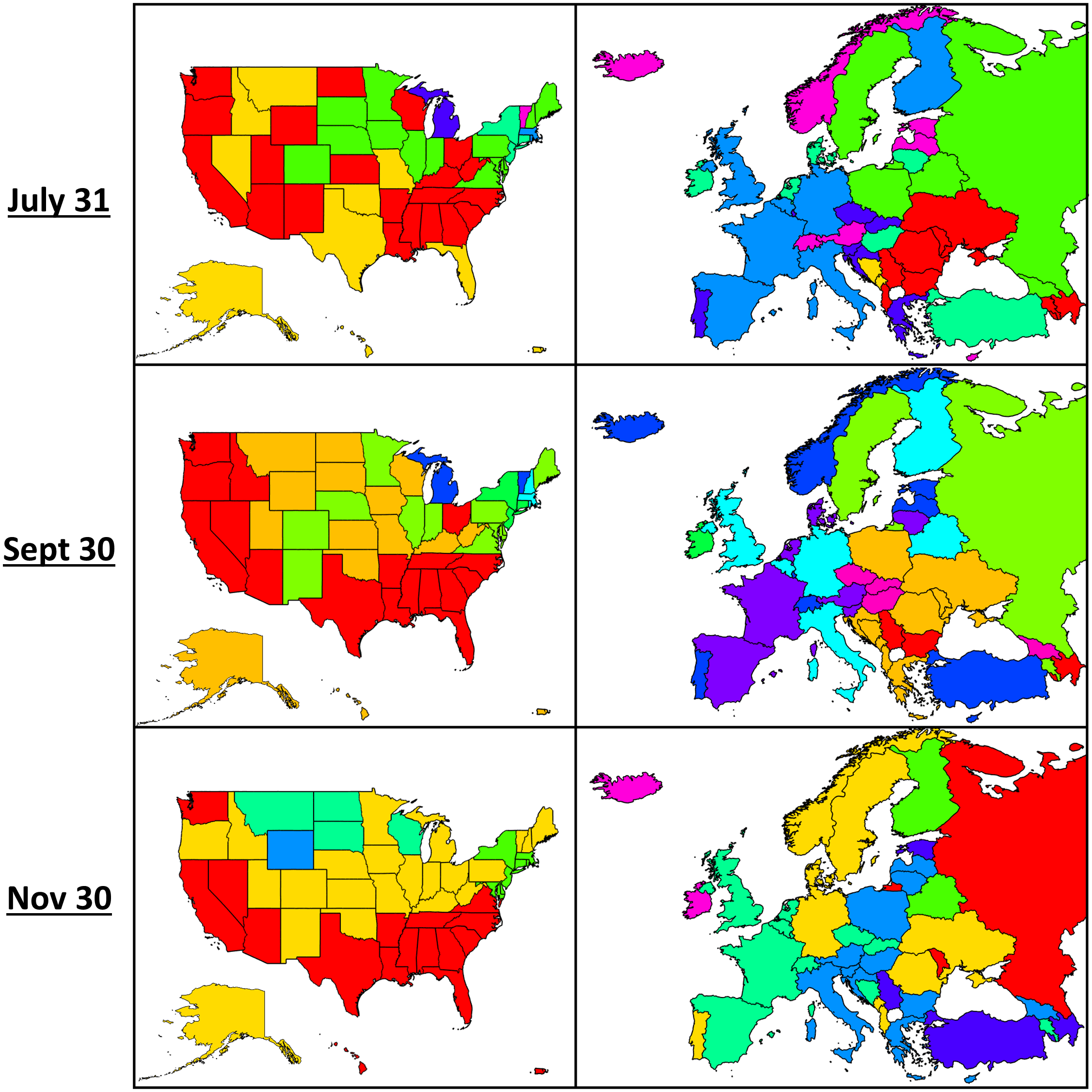} 
    \includegraphics[height=2in]{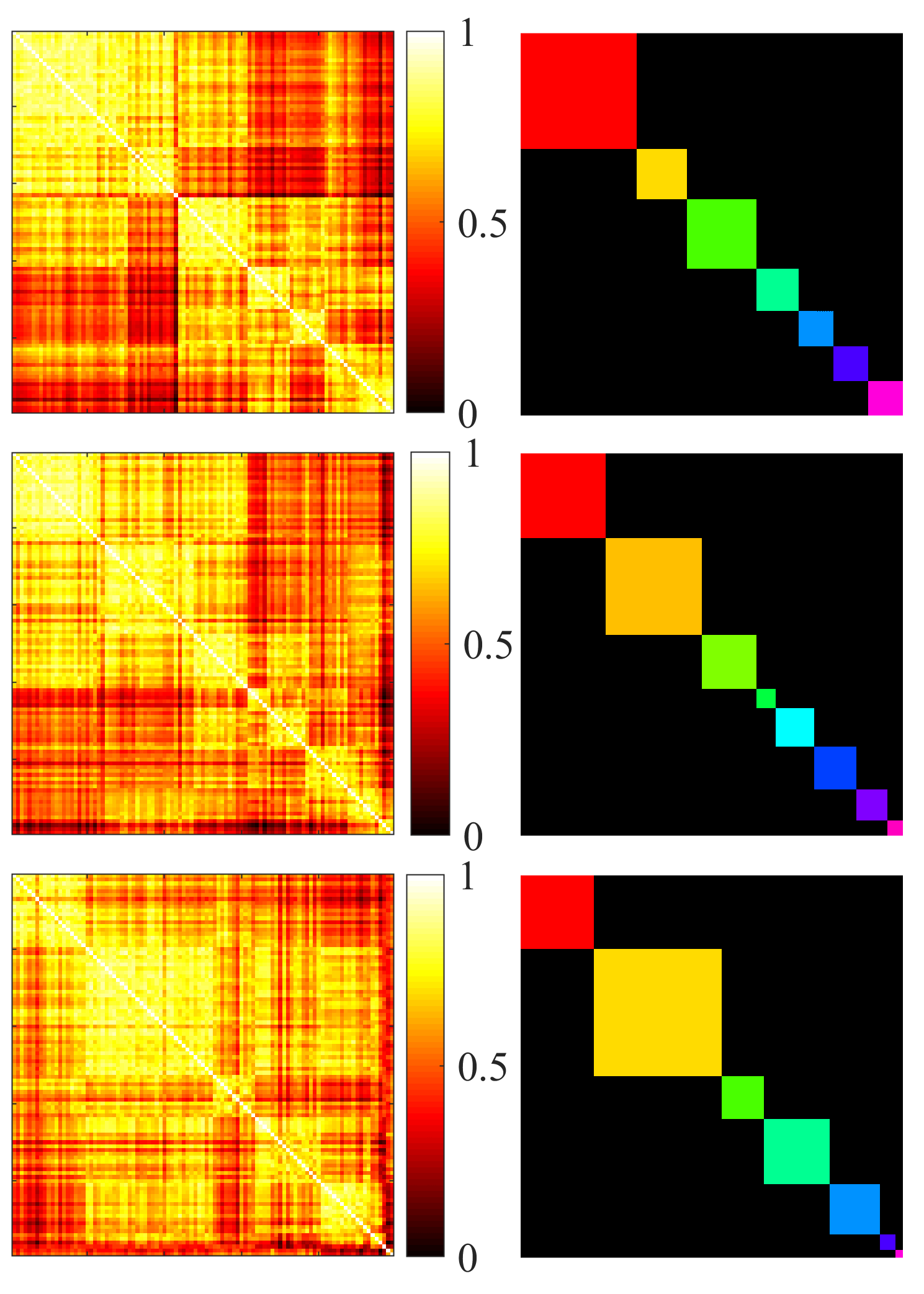} \\
    \includegraphics[width=\linewidth]{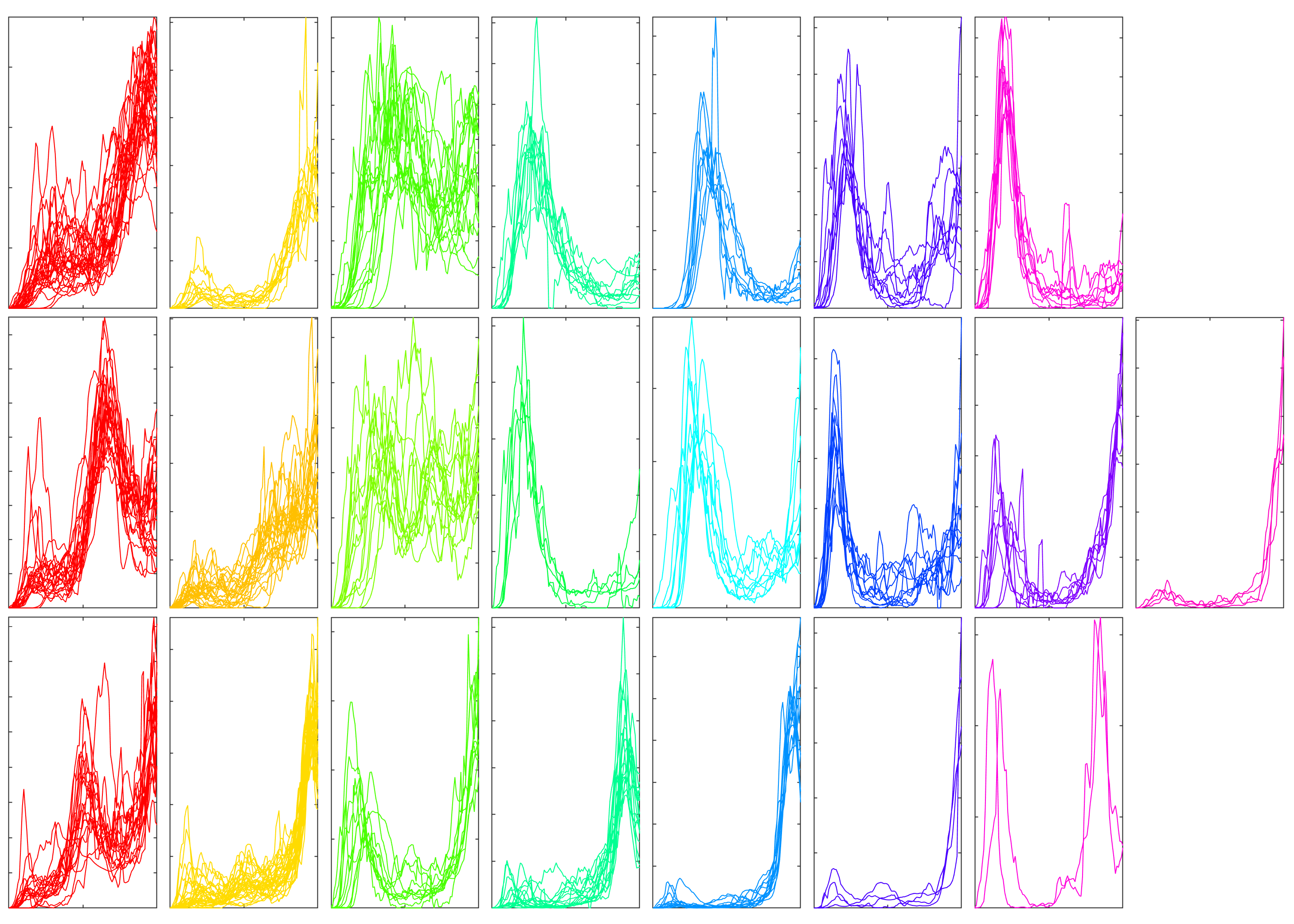} \\
    \includegraphics[width=\linewidth]{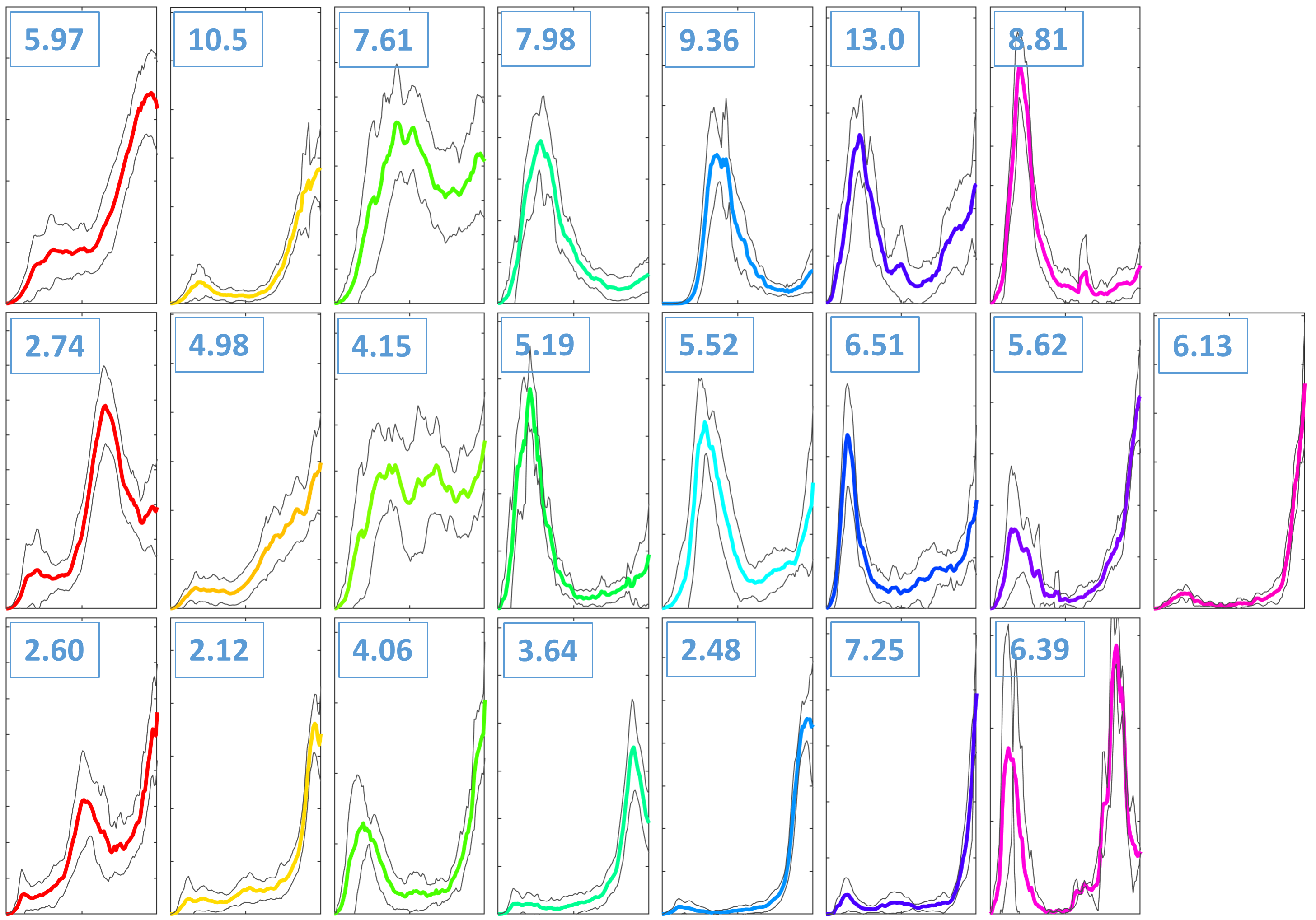} 
    \caption{Clustering results on the COVID-19 data set for the $\ltwo$ metric, method (a). Each group of three consecutive rows of image panels shows results gathered up to dates July 31, September 30, and November 30, respectively. That is, rows 1, 4, and 7 correspond to the July 31 dataset; rows 2, 5, and 8 correspond to the September 30 dataset; and rows 3, 6, and 9 to the November 30 dataset. From top to bottom, the first three rows show a US state \& European country map colorized according to class label, the pairwise similarity matrix, and the pairwise class inclusion matrix. The next three rows show the corresponding clusters, and the final three rows show the corresponding cluster cross-sectional means $\pm$ one cross-sectional standard deviation. The number in the upper-left corner of each panel here represents the average cross-sectional variance for that cluster ($\times 10^{-6}$).}
    \label{fig:covid_clusters_L2}
\end{figure}

\begin{figure}[ht]
    \centering
    \includegraphics[height=2in]{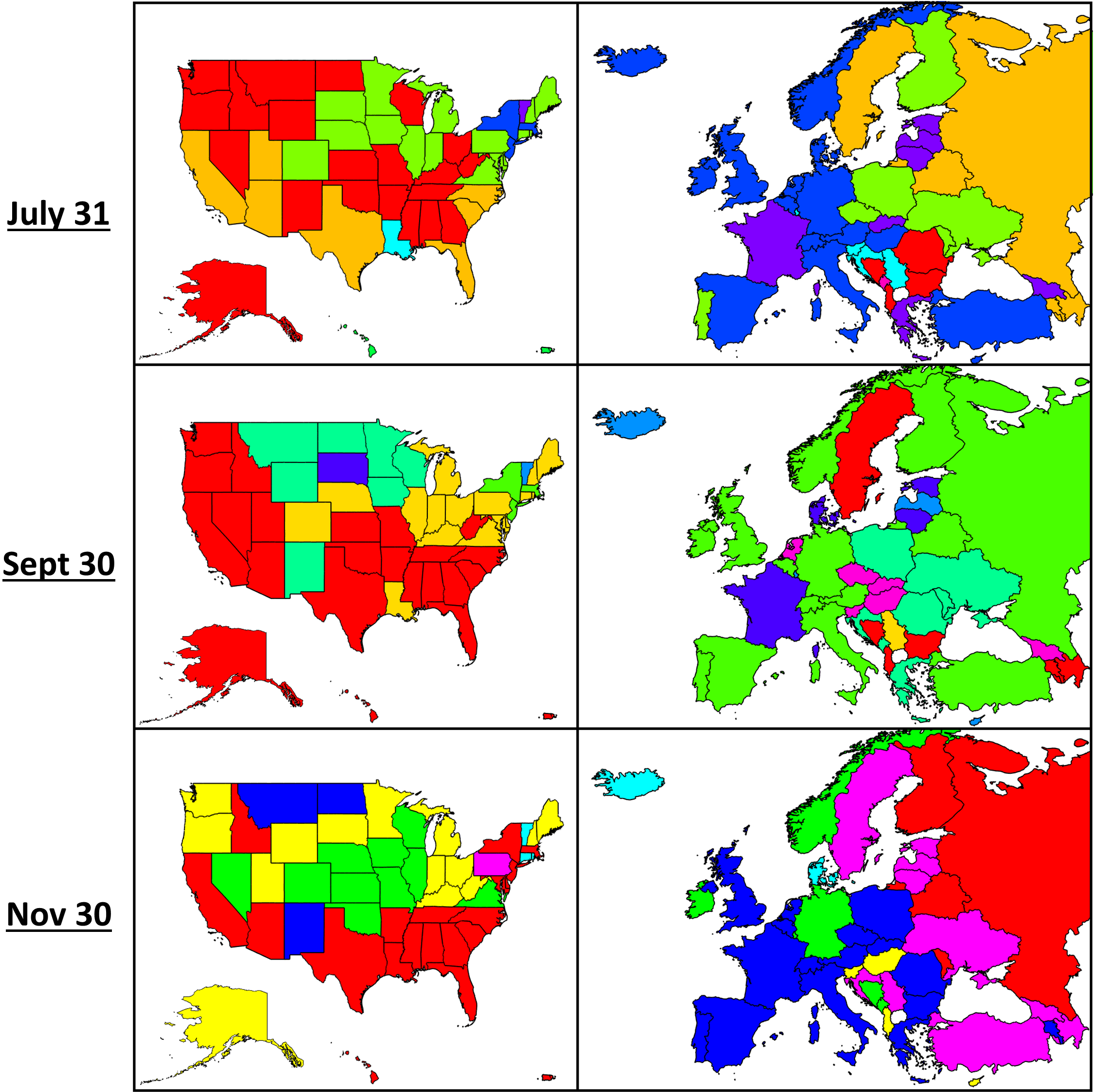} 
    \includegraphics[height=2in]{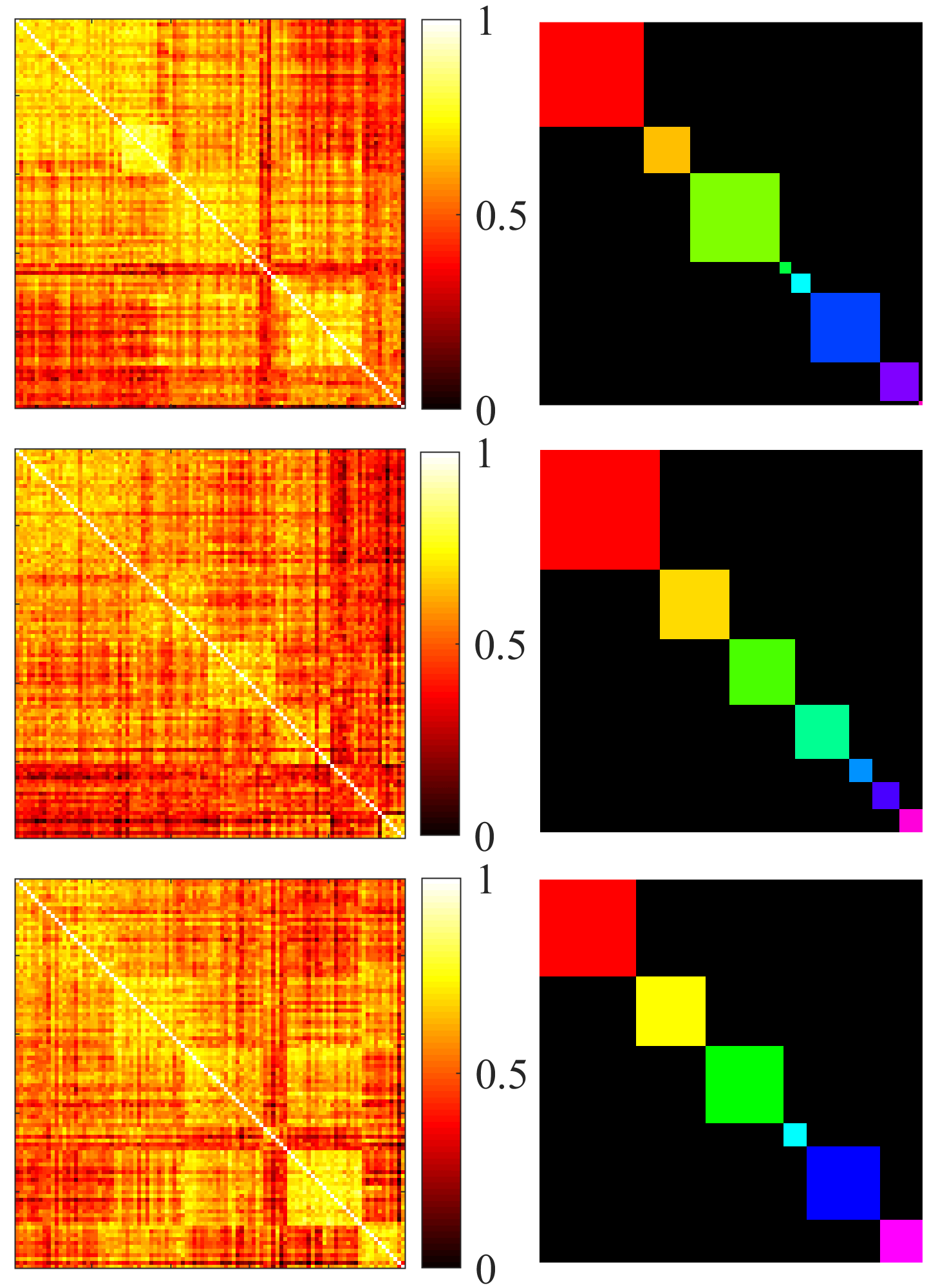} \\
    \includegraphics[width=\linewidth]{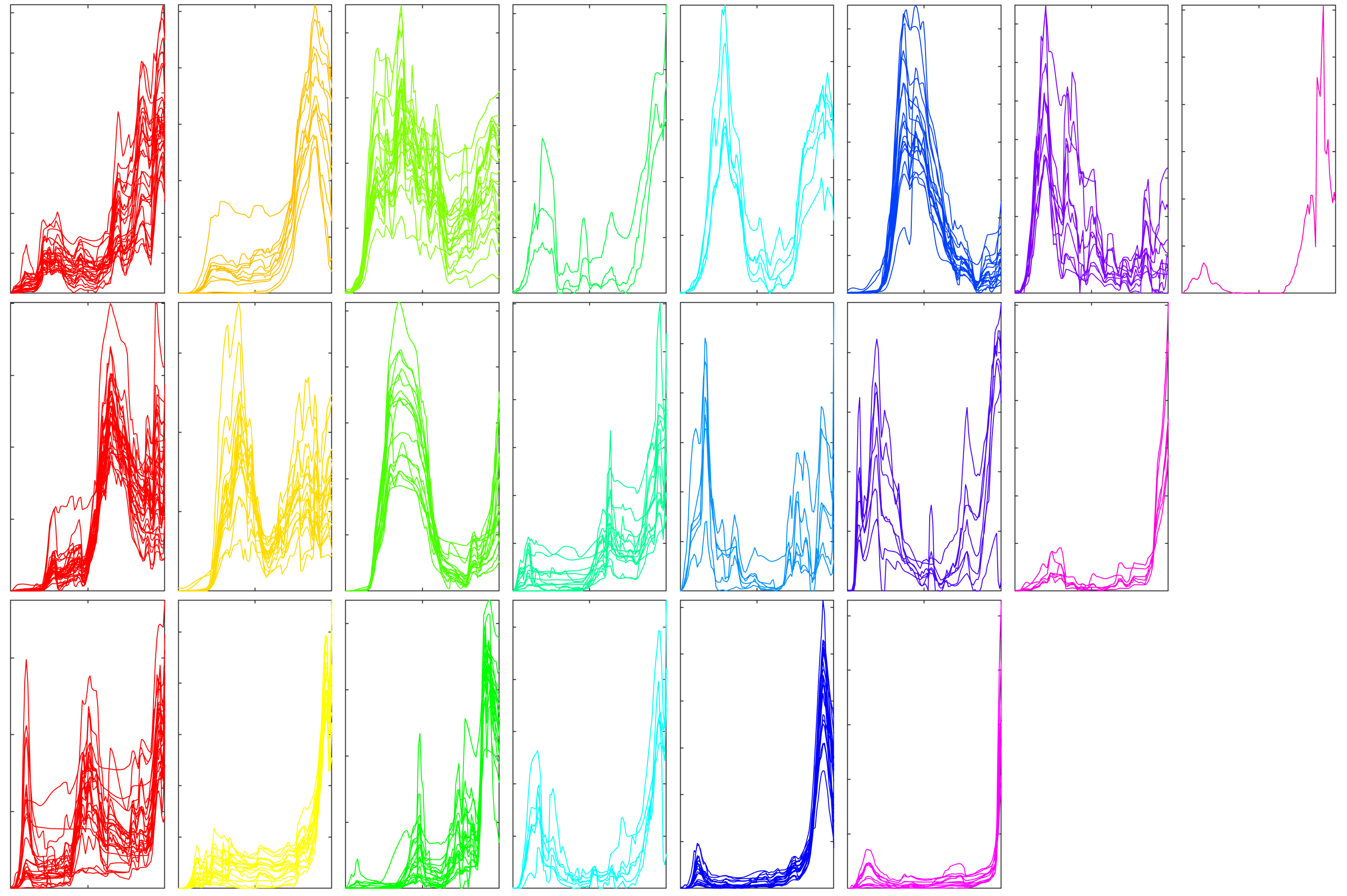} \\ 
    \includegraphics[width=\linewidth]{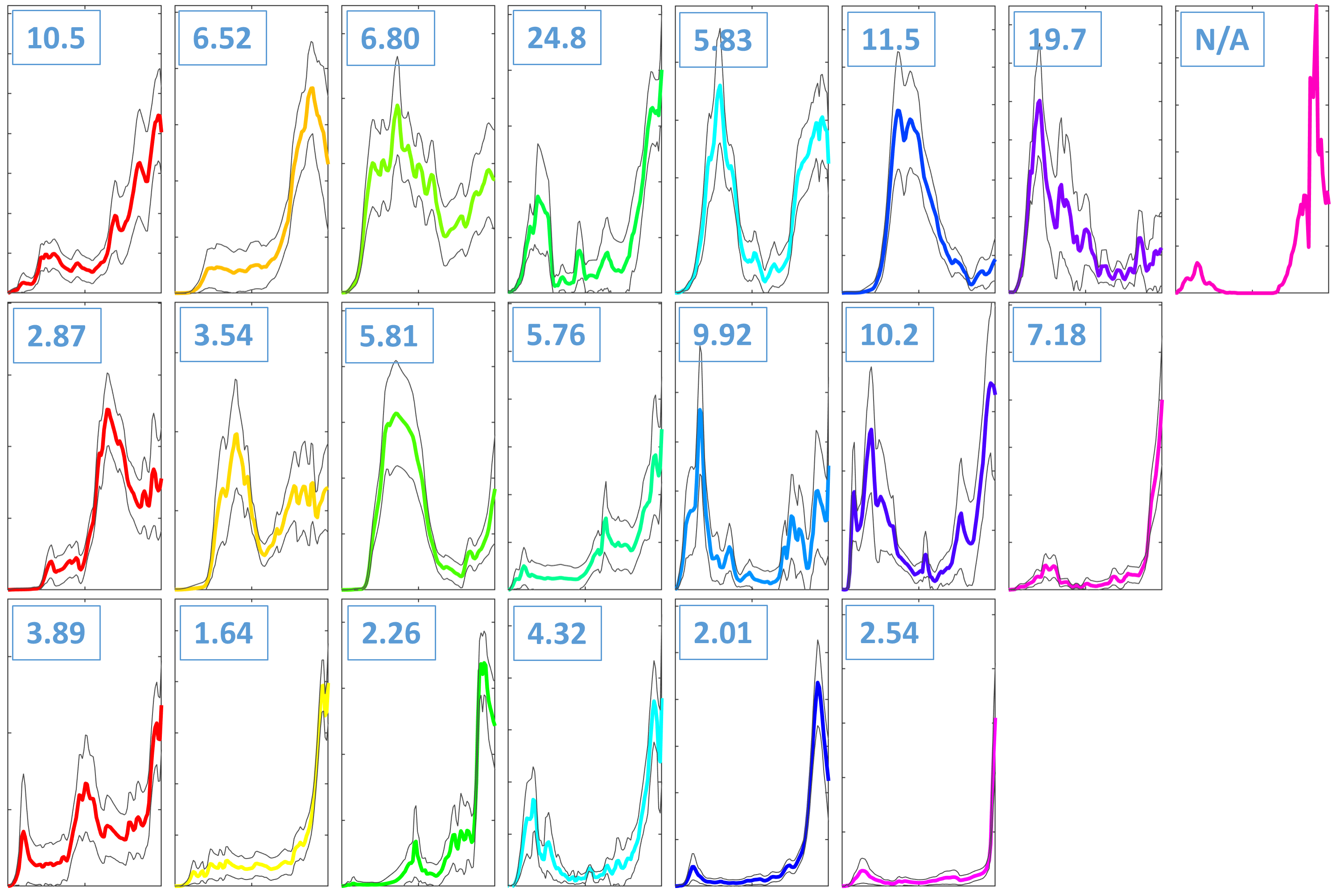} 
    \caption{Clustering results on the COVID-19 data set for the elastic metric with fixed endpoints, method (b). The figure description is the same as that of Fig.\ \ref{fig:covid_clusters_L2}.}
    \label{fig:covid_clusters_elastic}
\end{figure}

\begin{figure}[ht]
    \centering
    \includegraphics[height=2in]{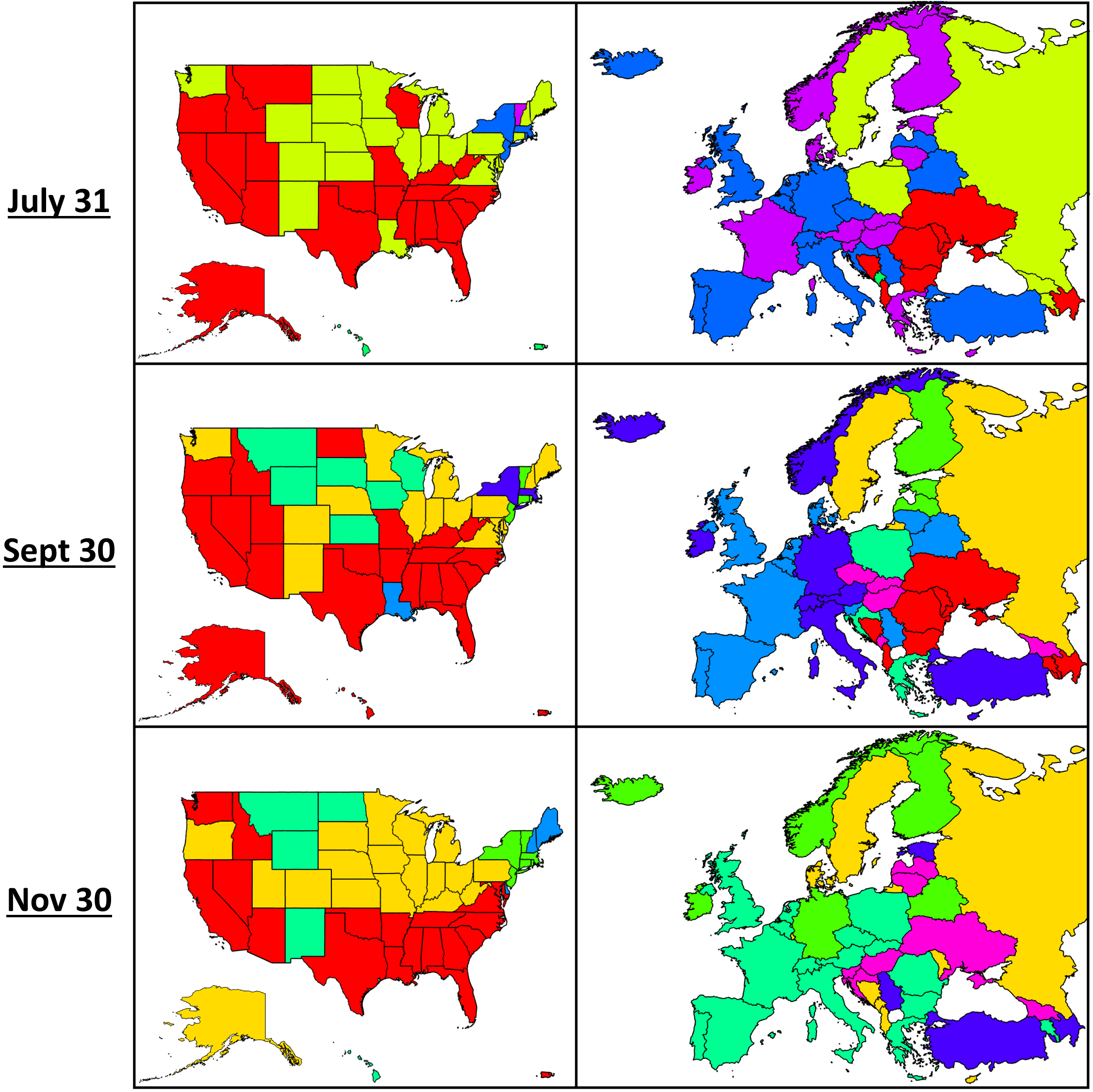} 
    \includegraphics[height=2in]{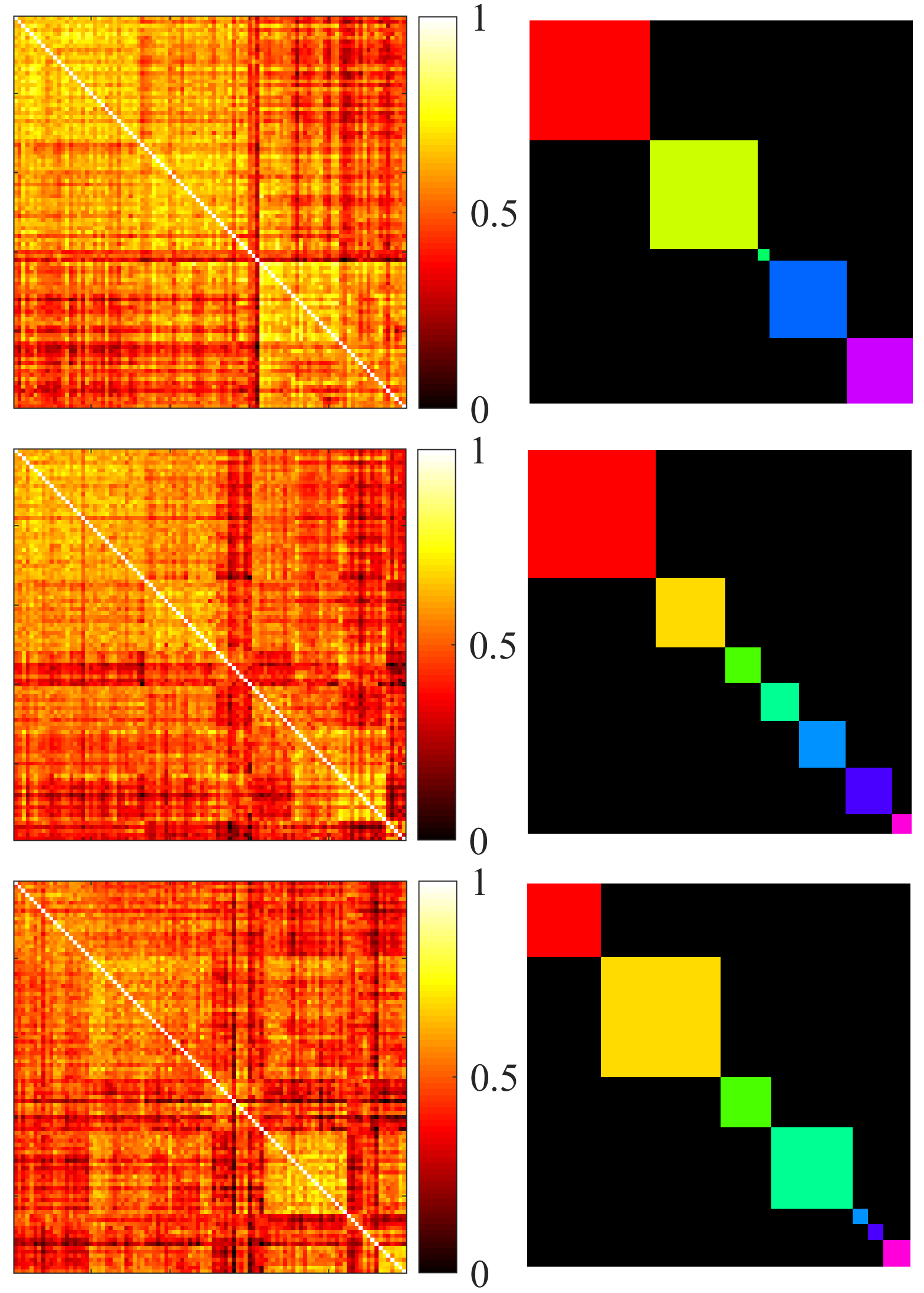} \\ 
    \includegraphics[width=\linewidth]{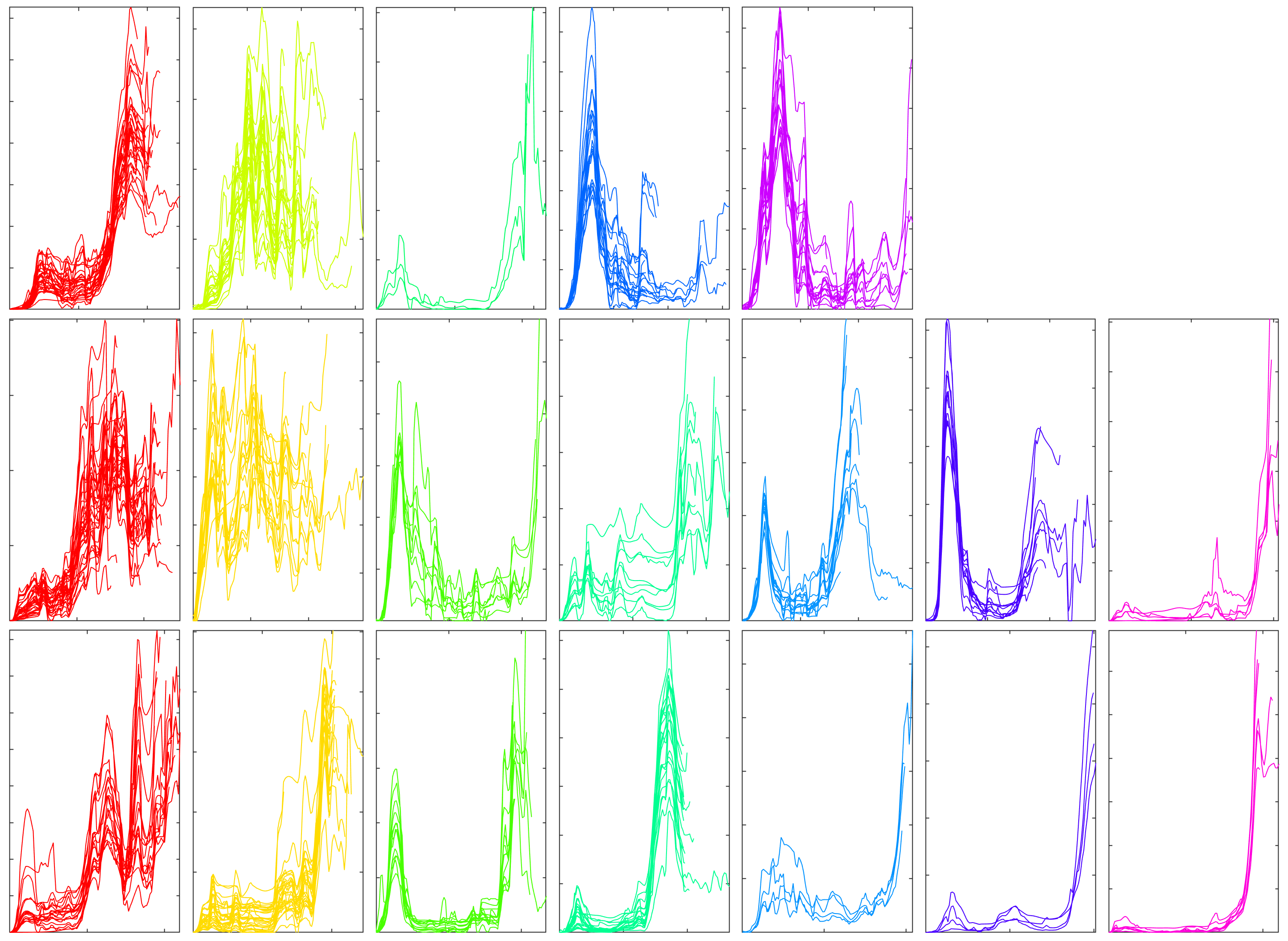} \\
    \includegraphics[width=\linewidth]{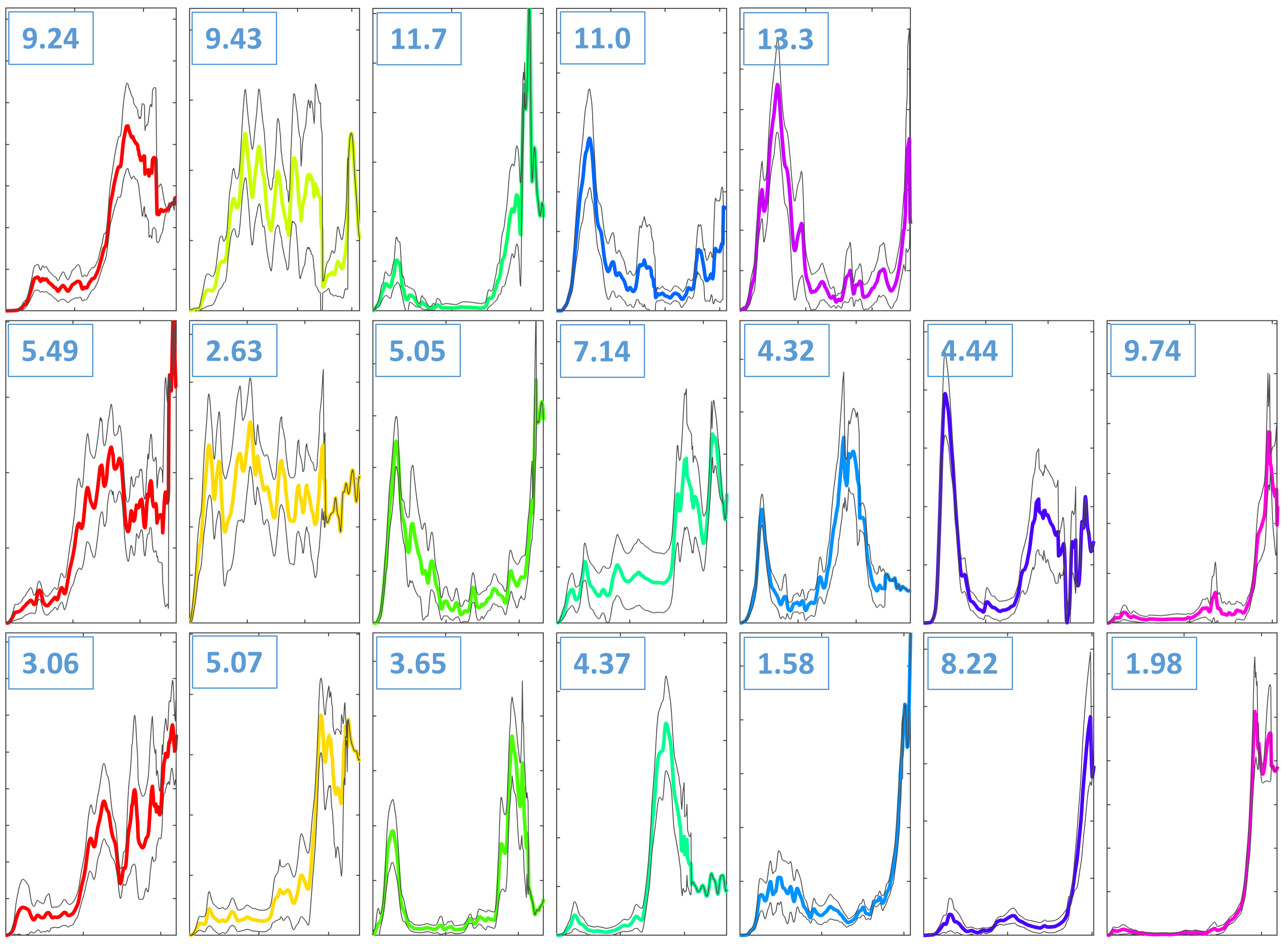} 
    \caption{Clustering results on the COVID-19 data set for the elastic metric with partial matching, method (c). The figure description is the same as that of Fig.\ \ref{fig:covid_clusters_L2}.}
    \label{fig:covid_clusters_partialElastic}
\end{figure}

\noindent
{\bf Dataset 2}:
Next, we apply the clustering algorithm on the real COVID-19 datasets. Figs.\ \ref{fig:covid_clusters_L2}, \ref{fig:covid_clusters_elastic}, and \ref{fig:covid_clusters_partialElastic} show the results of the Bayesian clustering using methods (a), (b), and (c), respectively. The organization of each of the three figures is the following. Each group of three consecutive rows of image panels shows results on the July 31, September 30, and November 30 datasets, respectively. The first group of three rows shows the color-coded country/state maps, the permuted pairwise similarity matrices, and the associated block-diagonal class inclusion matrices with the same color scheme as their associated maps. The next group of three rows shows the resulting function clusters, and the final group of three rows shows the cluster cross-sectional means with cross-sectional standard error bounds. Similar to Fig.\ \ref{fig:sim_clusters} with methods (b) and (c), the cluster members in Figs.\ \ref{fig:covid_clusters_elastic} and \ref{fig:covid_clusters_partialElastic} are mutually aligned within clusters in the tiled cluster display. Also similarly, the vertical axis has been scaled in each cluster tile so that the maximum value is equal to the maximum function value in each cluster set.    

There are quite a few inferences one can draw from this set of COVID-19 results. Firstly, and perhaps most notably, there is a high degree of geographical correlation of cluster members; i.e.\ neighboring states/countries are typically more likely to be of the same cluster member than not. 
Although it seems as though this geographic correlation is higher with our method (c), there are clusters present in all 9 analyses that consist of only European countries and other clusters that consist of mostly US states. A second clear inference one can draw is that method (c) yields less clusters than the other two methods on average due to its relative invariance to partial observation. 
Since method (c) tends to eliminate clusters with redundant shapes, it preserves true virus trajectory features, and groups states/countries more accurately compared to methods (a) and (b). Noteworthy shape-based features that discriminate between clusters present themselves as the number and relative intensity of waves of virus transmission. For example, one cluster could exhibit a small hump followed by a rapidly rising large hump, or, in the case of the July 31 data, another cluster could exhibit an initial large hump and a subsequent tapering off with little or no second wave. Finally, we can see that method (c) yields the lowest within-class cross-sectional variance on average per cluster, suggesting that these clusters are the tightest overall of the three methods. The mean value over all clusters for each of the three methods are the following: (a) $6.03 \times 10^{-6}$, (b) $7.38 \times 10^{-6}$, and (c) $5.82 \times 10^{-6}$.

Another important observation that we can make, and support through quantitative evidence, is that the similarity matrices in method (c) have less prominent block diagonal structures than that of methods (a) and (b). That is, although they may appear distinct and different to the naked eye, individual clusters are in reality more similar to each other with respect to method (c) than with other methods. This phenomenon is in stark contrast to that of the simulated data, where clusters were very distinct and separable both visually and with respect to the similarity measure itself. With the COVID data, it seems as though a shape model that is invariant to the large diffeomorphism space of time-warping and time-stretching allows for a homogenization of infection rate shapes in function space. In other words, save for a few outliers, this analysis shows that the states and countries all have more similar virus trajectory shapes compared to the other methods.

\begin{figure}[ht]
    \centering
    \includegraphics[width=\linewidth]{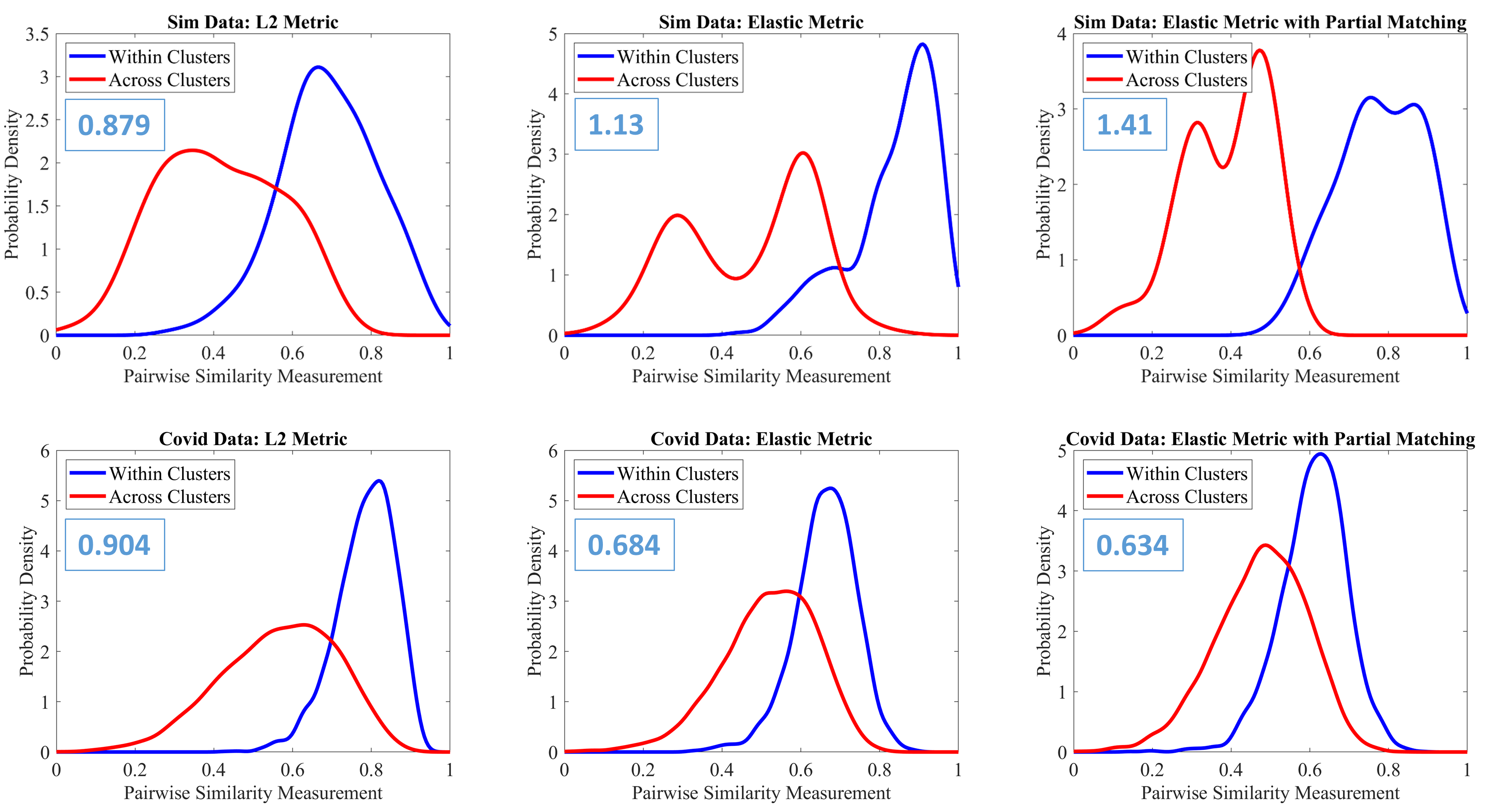}
    \caption{Estimated probability density functions of within-class and across-class pairwise similarity values. The top row corresponds to the simulated dataset, and the bottom row corresponds to the real COVID-19 datasets. The rows from left to right correspond to methods (a), (b), and (c) respectively. The number value in the top left of each panel indicates the distance measure, or separability, of each pair of density functions.}
    \label{fig:separability}
\end{figure}

Fig.\ \ref{fig:separability} provides evidence of the above claim. Here, we provide a measure of separability between within-class similarity values and across-class similarity values for each of the three methods and for both simulated and real datasets. Within-class similarity values are those that lie within the block diagonal structures on the upper triangular portion of the similarity matrix, and the across-class similarity values are those that lie outside the block diagonal structures. Each panel of Fig.\ \ref{fig:separability} shows the kernel density estimated probability density functions for within-class similarity values (blue) and across-class similarity values (red) for that particular dataset and method. The top row shows the results for the simulated data, and the bottom row shows results for the COVID data. The columns left to right represent results for methods (a), (b), and (c) respectively. The number shown in the box underneath the legend is our measure of separability, which is computed as a distance measure between the two probability distributions. We use the arclength between half-densities as our distance, or separability, formula (see \cite{srivastava-klassen:2016}), given by $\cos^{-1}(\langle \sqrt{f}_{in}, \sqrt{f}_{out} \rangle)$, where $f_{in}$ is the estimated within-class similarity value density and $f_{out}$ is the estimated across-class similarity value density. According to Fig.\ \ref{fig:separability}, the separability measure for the simulated data increases from left to right, with values of $0.879$, $1.13$, and $1.41$ for methods (a), (b), and (c), respectively. Visually, one can see that the red and the blue densities become less overlapped as we move from the left to right panels. Contrarily with the COVID data, the separability values decrease from left to right, with values of $0.875$, $0.695$, and $0.626$ for methods (a), (b), and (c), respectively. Visually, one can see that the red and blue densities overlap more and become more similar as we move from the left to right panels. This phenomenon provides evidence that states and countries have overall more similar virus trajectories with respect to method (c) than the other methods. Since the opposite effect is observed in simulated data, this phenomenon is due to the input data and not an artifact of the methodology used.

\section{Summary \& Future Work}

This paper presents a novel methodology to remove an important limitation of most past works in elastic matching, comparisons, and analysis of functional data, namely that the boundaries are fixed and registered. In addition to exhibiting phase variability within the observation interval, real data often also exhibits boundary variability, and the past methods fail to account for such boundary issues. The proposed elastic partial matching method forms a joint action of time-warping and time-scaling groups and searches for an optimal nonlinear warping along with a sliding right boundary to best match functions. As seen through experiments on simulated and COVID-19 rate curves, this additional freedom allows for more natural alignments and tighter, more visually distinct clusters compared to past methods.

Although the use of Riemannian elastic framework was motivated from a statistical perspective, the subsequent statistical tools are not developed in this paper due to a lack of space. The shape distance $d_s$ defined here is a proper distance on the quotient space ${\cal F}_0/G$ and can be used to define sample means, sample covariance, and tangent PCA based statistical models. These models can be used further for testing and classification of functional data.  

This framework can easily be applied to the problem of partially matching shapes of curves in Euclidean spaces~\cite{sebastian_kimia:03,partial-shape:2018,mahesh:2011,shilane:2006}. Very often, planar curves extracted from image data suffer from partial obscuration and missing parts. The proposed elastic partial framework naturally extends to curves in $\real^n$ and can be very useful in matching partially observed shapes in that context. 
Another important application of our elastic partial matching framework is human activity analysis~\cite{ashok-srivastava-etal-TIP:09,benamor-etal-PAMI:2015,anirudh-PAMI:2017}. As stated in these papers, classification of observed activities requires temporal synchronization, modulo variable execution rates, and this matching process can benefit from flexible boundaries. 

\section*{Acknowledgements}
This research was supported in part by the grants NSF CDS\&E DMS 1621787, NSF CIF 1617397, and NSF CDS\&E DMS 1953087 to AS; and by the Naval Innovative Science and Engineering (NISE) program at the Naval Surface Warfare Center Panama City Division (NSWC PCD) to DB. 

\appendices


\section{Proof of Lemma 4}

According to the Lemma, for $(c_1,q_1),(c_2,q_2)\in\ltwo_0$ and $b=b(\xi)=\min(c_1,c_2e^{-\xi})$, define the function $\tilde{q}_2$ as 
$$
\tilde{q}_2(t)=\left\{\begin{array}{ll}
    q_2(e^{\xi}b(\xi)\int_0^{t/b(\xi)}\psi(s)^2ds)e^{\xi/2}\psi(t/b(\xi)) & t\leq b(\xi) \\
    q_2(e^{\xi}t)e^{\xi/2} & t>b(\xi) 
\end{array}\right.
$$
First, we start by computing $b'(\xi)=-c_2e^{-\xi}1_{[\xi\geq -\log(c_1/c_2)]}$. Now, we can compute the derivative of $\tilde{q}_2$ with respect to $\xi$ when $t\leq b(\xi)$, and later we will compute the derivative for the case $t>b(\xi)$. For $t\leq b(\xi)$, we have that 
\begin{align*}
    \frac{\partial\tilde{q}_2}{\partial\xi} &= (e^{\xi}b\int_0^{t/b}\psi(s)^2ds \\
    &+ e^{\xi}(b'(\xi)\int_0^{t/b}\psi(s)^2ds-(t(b'(\xi)/b)\psi(t/b)^2)) \\
    &\cdot \; \dot{q}_2(e^{\xi}b\int_0^{t/b}\psi(s)^2ds)e^{\xi/2}\psi(t/b) \\
    &+ q_2(e^{\xi}b\int_0^{t/b}\psi(s)^2ds)(\frac{1}{2}e^{\xi/2}\psi(t/b) \\
    &- e^{\xi/2}(tb'(\xi)/b^2)\dot{\psi}(t/b)).
\end{align*}
Evaluated at $(\xi,\psi)=(\xi_{id},\psi_{id})=(0,1)$, and using the fact that $\int_0^{t/b(0)}\psi_{id}(s)^2ds=t/b(0)$, the above expression is equal to
\begin{align*}
    \frac{\partial\tilde{q}_2}{\partial\xi}(0,1) &= (b(0)t/b(0)-b'(0)t/b(0)+tb'(0)/b(0))\dot{q}_2(t) \\
    &+ q_2(t)(\frac{1}{2}+(tb'(0)/b(0)^2)\cdot 0) \\
    &= t\dot{q}_2(t)+\frac{1}{2}q_2(t).
\end{align*}
For the case of $t>b(\xi)$, we have that 
$$
\frac{\partial\tilde{q}_2}{\partial\xi} = e^{\xi}t\dot{q}_2(e^{\xi}t)e^{\xi/2}+q_2(e^{\xi}t)\frac{1}{2}e^{\xi/2}.
$$
Evaluated at $\xi=\xi_{id}=0$, the above expression is equal to
$$
\frac{\partial\tilde{q}_2}{\partial\xi}(0,1)=t\dot{q}_2(t)+\frac{1}{2}q_2(t).
$$
Thus, since the expression is the same for both cases $t\leq b(\xi)$ and $t>b(\xi)$, the above expression holds for all $t\geq 0$. $\Box$

\section{Proof of Lemma 5}

According to the Lemma, for $(c_1,q_1),(c_2,q_2)\in\ltwo_0$ and $b=\min(c_1,c_2)$, define the function $F:\Psi\to\real$ as
$$
F(\psi)=\int_0^b(q_1(t)-q_2(b\int_0^{t/b}\psi(s)^2ds)\psi(t/b))^2dt.
$$
In order to compute the gradient of this function at identity, we first need to write an expression for the directional derivative in the direction of a tangent vector $z\in T_{\psi_{id}}(\Psi)$. Using a variational approach, define a path $\alpha:(-\epsilon,\epsilon)\to\Psi$ such that $\alpha(0)=\psi_{id}$ and $\frac{\partial\alpha}{\partial\tau}(0)=z$. The directional derivative of $F$ at identity in the direction of $z$ is defined as $D_zF(\psi_{id})=\frac{\partial}{\partial\tau}F(\alpha(\tau))|_{\tau=0}$. Before computing the full directional derivative, we first define the function 
$$
\tilde{q}_2=q_2(b\int_0^{t/b}\alpha(\tau)(s)^2~ds) \alpha(\tau)(t/b)\ ,
$$ 
and compute the derivative of $\tilde{q}_2$ with respect to $\tau$ as
\begin{align*}
    \frac{\partial\tilde{q}_2}{\partial\tau} &= (b\int_0^{t/b}2\dot{\alpha}(\tau)\alpha(\tau)ds) \dot{q}_2(b\int_0^t\alpha(\tau)^2ds)\alpha(\tau)(t/b) \\
    &+ q_2(b\int_0^{t/b}\alpha(\tau)^2ds)\dot{\alpha}(\tau)(t/b).
\end{align*}
If we define the variable change $x=t/b$ and the function $\tilde{z}(x)=\int_0^xz(s)ds$, then the above expression evaluated at $\tau=0$ becomes
$$
\frac{\partial\tilde{q}_2}{\partial\tau}|_{\tau=0}=2b\tilde{z}(x)\dot{q}_2(bx)+q_2(bx)c(x).
$$
Taking into account this change of variables, the full derivative of $F$ with respect to $\tau$ and evaluated at $\tau=0$ is given by 
\begin{align*}
    D_zF(\psi_{id}) &= -2\int_0^1(q_1(bx)-q_2(bx))\frac{\partial\tilde{q}_2}{\partial\tau}|_{\tau=0}bdx \\
    &= -2b\int_0^1(q_1(bx)-q_2(bx))(2b\tilde{z}(x)\dot{q}_2(bx) \\
    &+ q_2(bx)z(x))dx.
\end{align*}

In order to compute the gradient of $F$, we must first express the directional derivative in the form $D_zF(\psi_{id})=\langle w,z \rangle$ for some function $w\in \ltwo([0,1],\real)$. In order to do so, we make use of integration by parts to rewrite the above integral via the following procedure. First rewrite the directional derivative as
\begin{align*}
    D_zF(\psi_{id}) &= -4b^2\int_0^1(q_1(bx)-q_2(bx))\dot{q}_2(bx)\tilde{z}(x)dx \\
    &- 2b\int_0^1(q_1(bx)-q_2(bx))q_2(bx)z(x)dx.
\end{align*}
Now, we can use integration by parts to rewrite the first integral above by setting $u(x)=\tilde{z}(x)$, $v'(x)=(q_1(bx)-q_2(bx))\dot{q}_2(bx)$, and then letting $\int_0^1u(x)v'(x)dx=[u(x)v(x)]_0^1-\int_0^1u'(x)v(x)dx$. Ultimately, this procedure yields
\begin{align*}
    & \int_0^1(q_1(bx)-q_2(bx))\dot{q}_2(bx)\tilde{z}(x)dx \\
    &= -\int_0^1(\int_0^x(q_1(bs)-q_2(bs))\dot{q}_2(bs)ds)z(x)dx
\end{align*}
Therefore, if we define the function $w$ as
\begin{align*}
    w(x) &= 4b^2\int_0^x(q_1(bs)-q_2(bs))\dot{q}_2(bs)ds \\
    &- 2b(q_1(bx)-q_2(bx))q_2(bx),
\end{align*}
then we can write the directional derivative in the form $D_zF(\psi_{id})=\langle w,z \rangle$. Finally, the Riemannian gradient of $F$ at identity is given by the projection of $w$ to the tangent space $T_{\psi_{id}}(\Psi)$. This projection, and thus the gradient, is computed as 
$$
\nabla F(\psi_{id})=w(x)-\int_0^1w(x)dx. \;\; \Box
$$

\bibliography{bibfile}

\begin{thebibliography}{10}

\bibitem{Alt95}
H.~Alt and M.~Godau.
\newblock Computing the {F}rechet distance between two polygonal curves.
\newblock {\em International Journal of Computational Geometry and
  Applications}, 5:75 -- 91, March 1995.

\bibitem{benamor-etal-PAMI:2015}
B.~Ben Amor, J.~Su, and A.~Srivastava.
\newblock Action recognition using rate-invariant analysis of skeletal shape
  trajectories.
\newblock {\em IEEE Transaction on Pattern Analysis and Machine Intelligence},
  38(1):1--13, 2016.

\bibitem{anirudh-PAMI:2017}
R.~Anirudh, P.~Turaga, J.~Su, and A.~Srivastava.
\newblock Elastic functional coding of riemannian trajectories.
\newblock {\em IEEE Transactions of Pattern Analysis and Machine Intelligence},
  39(5):922--936, 2017.

\bibitem{Belkis-2018}
A.~Belkis, J.~Demongeot, A.~Laksaci, and M.~Rachdi.
\newblock Functional data analysis: estimation of the relative error in
  functional regression under random left-truncation.
\newblock {\em Journal of Nonparametric Statistics}, 30(2):472--490, 2018.

\bibitem{buchin09}
K.~Buchin, M.~Buchin, and Y.~Wang.
\newblock Exact algorithms for partial curve matching via the {F}rechet
  distance.
\newblock In {\em Proceedings of the 2009 Annual ACM-SIAM Symposium on Discrete
  Algorithms}, pages 645 -- 654, January 2009.

\bibitem{cui-2009}
M.~Cui, J.~Femiani, J.~Hu, P.~Wonka, and A.~Razdan.
\newblock Curve matching for open 2{D} curves.
\newblock {\em Pattern Recognition Letters}, 30:1--10, 2009.

\bibitem{Delaigle-Hall-2013}
A.~Delaigle and P.~Hall.
\newblock Classification using censored functional data.
\newblock {\em Journal of the American Statistical Association},
  108(504):1269--1283, 2013.

\bibitem{elmi-2009}
A.F. Elmi.
\newblock {\em Curve registration in functional data analysis with
  informatively censored event-times}.
\newblock PhD thesis, Univ. of Penn., 2009.

\bibitem{Ferraty2006-sp}
F.~Ferraty and P.~Vieu.
\newblock {\em Nonparametric Functional Data Analysis: Theory and Practice}.
\newblock Springer Series in Statistics. Springer New York, 2006.

\bibitem{shilane:2006}
T.~Funkhouser and P.~Shilane.
\newblock Partial matching of 3{D} shapes with priority-driven search.
\newblock In {\em Eurographics Symposium on Geometry Processing}, 2006.

\bibitem{hsing-eubank:2015}
T.~Hsing and R.~Eubank.
\newblock {\em Theoretical Foundations of Functional Data Analysis, with an
  Introduction to Linear Operators}.
\newblock Wiley, 2015.

\bibitem{huang-BFGS-elastic}
W.~Huang, K.A. Gallivan, A.~Srivastava, and P.-A. Absil.
\newblock Riemannian optimization for registration of curves in elastic shape
  analysis.
\newblock {\em Journal of Mathematical Imaging and Vision}, 54(3):320--343,
  2016.

\bibitem{survival-klein}
J.P. Klein and M.L. Moeschberger.
\newblock {\em Survival Analysis: Techniques for Censored and Truncated Data}.
\newblock Springer Verlag New York, 2003.

\bibitem{kneip-ramsay:2008}
A.~Kneip and J.O. Ramsay.
\newblock Combining registration and fitting for functional models.
\newblock {\em Journal of the American Statistical Association}, 103(483),
  2008.

\bibitem{fda-reimharr}
P.~Kokoszka and M.~Reimherr.
\newblock {\em Introduction to Functional Data Analysis}.
\newblock Chapman and Hall, 2017.

\bibitem{dehan-2018}
D.~Kong, J.G. Ibrahim, E.~Lee, and H.~Zhu.
\newblock {FLCRM}: Functional linear cox regression model.
\newblock {\em Biometrics}, 74(1):109--117, 2018.

\bibitem{jung-joint-PA}
S.~Lee and S.~Jung.
\newblock Combined analysis of amplitude and phase variations in functional
  data.
\newblock {\em arXiv:1603.01775v2}, 2017.

\bibitem{mahesh:2011}
A.~Maheshwari, J.R. Sack, K.~Shahbaz, and H.~Zarrabi-Zadeh.
\newblock Improved algorithms for partial curve matching.
\newblock In {\em Algorithms – ESA 2011. Lecture Notes in Computer Science,
  vol 6942}. Springer, Berlin, Heidelberg, 2011.

\bibitem{marron-etal-EJS:2014}
J.S. Marron, J.O. Ramsay, L.M. Sangalli, and A.~Srivastava.
\newblock Statistics of time warpings and phase variations.
\newblock {\em Electronic Journal of Statistics}, 8(2):1697--1702, 2014.

\bibitem{marron-etal-EJS:2015}
J.S. Marron, J.O. Ramsay, L.M. Sangalli, and A.~Srivastava.
\newblock Functional data analysis of amplitude and phase variation.
\newblock {\em Statistical Science}, 30(4):468--484, November 2015.

\bibitem{McBride03}
J.C. McBride and B.B. Kimia.
\newblock Archaeological fragment reconstruction using curve-matching.
\newblock In {\em 2003 Conference on Computer Vision and Pattern Recognition
  Workshop}, volume~1, 2003.

\bibitem{ramsay-gasser-94}
J.O. Ramsay, R.D. Bock, and T.~Gasser.
\newblock Comparison of height acceleration curves in the fels, zurich, and
  berkeley growth data.
\newblock {\em Annals of Human Biology}, 22(5):413--426, 1999.

\bibitem{ramsay-li-RSSB:98}
J.O. Ramsay and X.~Li.
\newblock Curve registration.
\newblock {\em Journal of the Royal Statistical Society, Ser. B}, 60:351--363,
  1998.

\bibitem{ramsay-silverman-2005}
J.O. Ramsay and B.W. Silverman.
\newblock {\em Functional Data Analysis, Second Edition}.
\newblock Springer Series in Statistics, 2005.

\bibitem{robinson-thesis}
D.~Robinson.
\newblock {\em Functional data analysis and partial shape matching in
  square-root velocity framework}.
\newblock FSU Dissertaion, 2012.

\bibitem{sebastian_kimia:03}
T.B. Sebastian, P.N. Klein, and B.B. Kimia.
\newblock On aligning curves.
\newblock {\em IEEE Transactions on Pattern Analysis and Machine Intelligence},
  25(1):116--125, 2003.

\bibitem{srivastava-klassen:2016}
A.~Srivastava and E.~Klassen.
\newblock {\em Functional and Shape Data Analysis}.
\newblock Springer Series in Statistics, 2016.

\bibitem{srivastava-etal-function:2011}
A.~Srivastava, W.~Wu, S.~Kurtek, E.~Klassen, and J.~S. Marron.
\newblock Registration of functional data using fisher-rao metric.
\newblock {\em arXiv}, arXiv:1103.3817, 2011.

\bibitem{Takagishi-Yadohisa}
M.~Takagishi and H.~Yadohisa.
\newblock Robust curve registration using the t distribution.
\newblock {\em Behaviormetrika}, 46:177–198, 2019.

\bibitem{tucker-et-al:2013}
J.D. Tucker, W.~Wu, and A.~Srivastava.
\newblock Generative models for functional data using phase and amplitude
  separation.
\newblock {\em Computational Statistics and Data Analysis}, 61:50--66, 2013.

\bibitem{ashok-srivastava-etal-TIP:09}
A.~Veeraraghavan, A.~Srivastava, A.K. Roy-Chowdhury, and R.~Chellappa.
\newblock Rate-invariant recognition of humans and their activities.
\newblock {\em IEEE Transactions on Image Processing}, 8(6):1326--1339, June
  2009.

\bibitem{partial-shape:2018}
C.~Yang, H.~Wei, and Q.~Yu.
\newblock A novel method for 2d nonrigid partial shape matching.
\newblock {\em Neurocomputing}, 275:1160--1176, 2018.

\bibitem{zhang2015}
Z.~Zhang, D.~Pati, and A.~Srivastava.
\newblock Bayesian clustering of shapes of curves.
\newblock {\em Journal of Statistical Planning and Inference}, 166:171 -- 186,
  2015.

\end{thebibliography}
\bibliographystyle{plain}

\begin{IEEEbiography}[{\includegraphics[height=1.25in,clip,keepaspectratio]{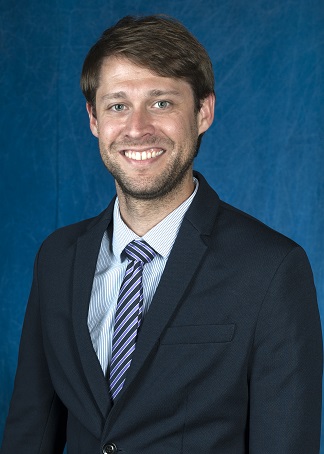}}]{Darshan Bryner}
Darshan Bryner is a research scientist and branch manager at the Naval Surface Warfare Center Panama City Division (NSWC PCD). He obtained his Ph.D. degree in Statistics from Florida State University in 2013, where his dissertation research focused on developing novel statistical shape models to improve image segmentation quality. As a research scientist at NSWC PCD, Dr. Bryner has conducted basic and applied research in the fields of statistical computer vision, functional data analysis, spatial statistics, optimization on nonlinear manifolds, and bioinformatics, leading to several publications in top-tier peer-reviewed journals and conferences. Since 2017, Dr. Bryner has served as the head of the Advanced Signal Processing and Automatic Target Recognition Branch (Code X23) at NSWC PCD.  
\end{IEEEbiography}

\begin{IEEEbiography}[{\includegraphics[width=1in,height=1.25in,clip,keepaspectratio]{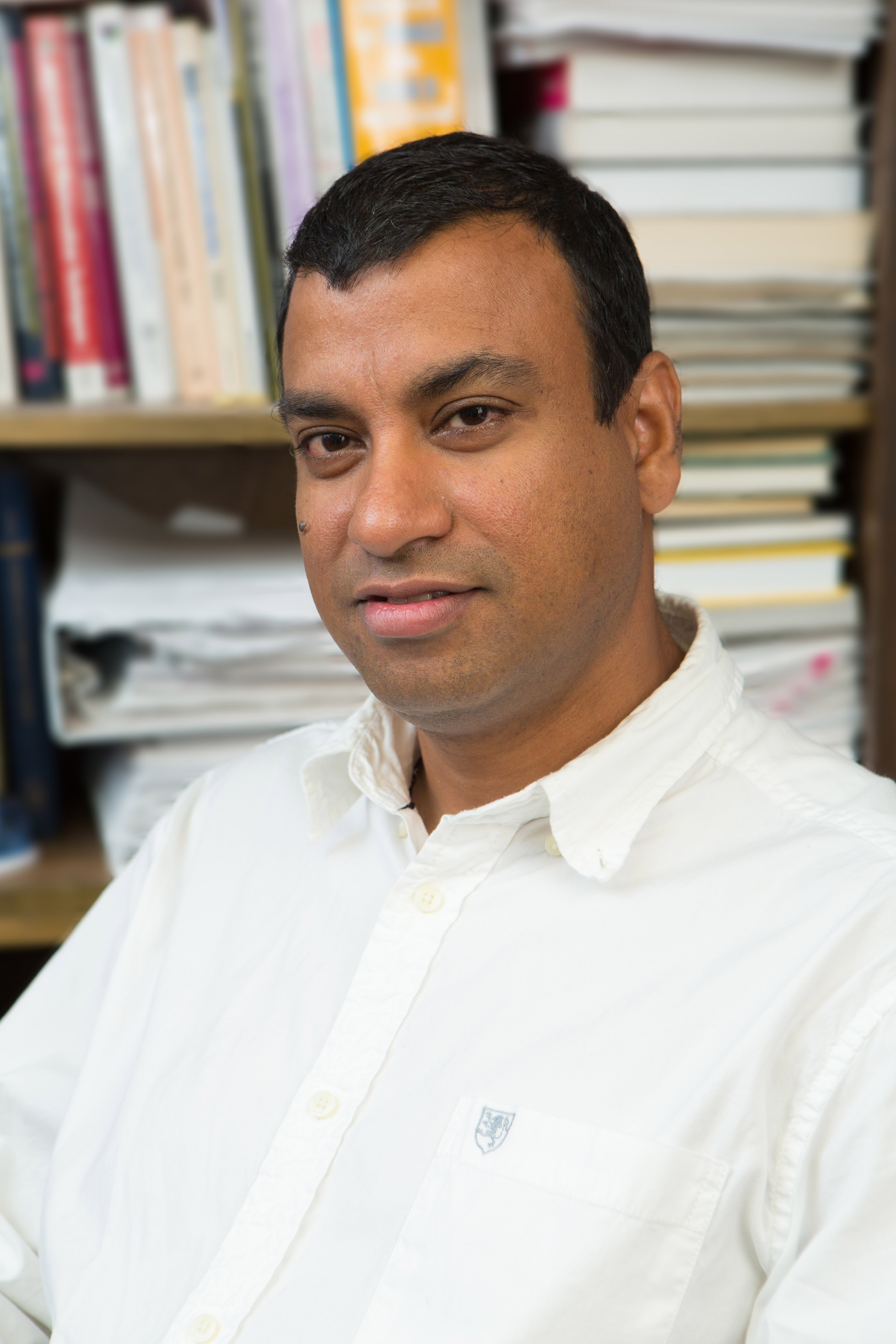}}]{Anuj Srivastava}
Anuj Srivastava is a Professor and a Distinguished Research Professor at the Florida State University.   His research interests include statistical analysis on nonlinear manifolds, statistical computer vision, functional data analysis, and shape analysis. He is a fellow of AAAS, IAPR, IEEE, and ASA. He has held several visiting positions at European universities, including INRIA, France;  the University of Lille, France; and Durham University, UK. He has coauthored more than 230 papers in peer-reviewed journals and top-tier conferences, and also several books, including the 2016 Springer textbook on "Functional and Shape Data Analysis". 
\end{IEEEbiography}

\end{document}